\documentclass[twocolumn]{aastex701}
\usepackage{amsmath}
\usepackage{enumitem}
\usepackage{xcolor}

\begin{document}

\title{How eccentric is the debris disk of $\epsilon$ Eridani? ALMA reveals a near-circular belt}

\author[orcid=0009-0006-8868-2932]{Oto Ulrich}
\affiliation{School of Physics, Trinity College Dublin, the University of Dublin, College Green, Dublin 2, Ireland}
\affiliation{Center for Astrophysics, Harvard \& Smithsonian, 60 Garden Street, Cambridge, MA 02138-1516, USA}
\email[show]{oto.ulrich@gmail.com}  

\author[orcid=0000-0002-4248-5443]{Joshua B. Lovell}
\affiliation{Center for Astrophysics, Harvard \& Smithsonian, 60 Garden Street, Cambridge, MA 02138-1516, USA}
\email[show]{joshualovellastro@gmail.com}  

\author[orcid=0000-0003-1526-7587]{David J. Wilner}
\affiliation{Center for Astrophysics, Harvard \& Smithsonian, 60 Garden Street, Cambridge, MA 02138-1516, USA}
\email{dwilner@cfa.harvard.edu}

\author[orcid=0000-0003-4623-1165]{Antranik A. Sefilian}
\affiliation{Department of Astronomy and Steward Observatory, University of Arizona, Tucson, AZ 85721, USA}
\email{asefilian@arizona.edu}

\author[orcid=0000-0001-8568-6336]{Mark Booth}
\affiliation{UK Astronomy Technology Centre, Royal Observatory Edinburgh, Blackford Hill, Edinburgh EH9 3HJ, UK}
\email{mark.booth@stfc.ac.uk}

\begin{abstract}

The nearby (3.2 pc) K2V star $\epsilon$~Eridani hosts the closest detected debris disk and an exo-Jupiter analog ($\epsilon$~Eridani~b).
In this work, we reanalyse ALMA Band~6 (1.29 mm) observations of $\epsilon$~Eridani to assess whether its outer debris belt exhibits a non-zero eccentricity, potentially induced by planet--disk interactions.
Considering two models, one that accounts for the disk's forced eccentricity ($e_f$) and another that accounts for a proper eccentricity $(e_p)$, we place stringent (99.7th-percentile) upper limits of $e_f < 2.6\%$ and $e_p < 6.8\%$, respectively, and find
tentative evidence for a non-zero forced eccentricity of $e_f {=} 1.62^{+0.39}_{-0.40}\%$ at the $4\sigma$ level.
This corresponds to the lowest millimeter-derived upper limit on the forced eccentricity of any debris disk to date, which, together with the low upper limit on the proper eccentricity, suggests that the outer belt is nearly circular and dynamically cold.
We further explore planetary system architectures that are consistent with these ALMA-based results in combination with Spitzer/IRAC, JWST/NIRCam F444W, and JWST/MIRI F2550W data.
Together, these analyses provide insight into potential planet--disk interactions in this nearby system, 
constraining any additional planet beyond $\epsilon$~Eridani~b to masses ranging from ${\sim} 2{\times}10^{-4}$ to $0.3\,M_{\rm Jup}$ at orbital separations of ${\sim}6{-}60$~au.
\end{abstract}

\keywords{\uat{Debris disks}{363} --- \uat{Circumstellar disks}{235} --- \uat{Eccentricity}{441} --- \uat{Planetary-disk interactions}{2204} --- \uat{Planetary system evolution}{2292}}

\section{Introduction} \label{sec:introduction}
Debris disks are belts of dust observed around main-sequence stars, produced by mutual planetesimal collisions, analogous to the Solar System's main asteroid and Kuiper belts \citep{Wyatt, Matthews_2014, Hughes_2018}.
Such collisions produce dust size distributions that span many orders of magnitude (from $\mu m$ to km), enabling studies of their properties across a range of wavelengths, either via scattered light or thermal emission \citep[see e.g. the optical, infrared, and millimeter surveys of][]{Bryden_2006, Eiroa_2013, Booth_2013, Holland_2017, Esposito_2020, Matra_2025, Engler_2025, ARKSI}. 

The morphology of debris disks can shed light on the underlying planetary system architecture, e.g. via planet--disk interactions.
For example, planets (if migrating) can trap planetesimals into mean motion resonances (MMRs) and sculpt clumpy structures \citep{Wyatt_2003}, sculpt the edges of debris disks via scattering and chaotic dynamics due to overlapping MMRs \citep{Quillen2006, Pearce_2024}, secularly perturb debris disks into eccentric and/or gapped morphologies \citep{Wyatt_1999, Yelverton2018, Sefilian2021, Sefilian2023}, initiate a collisional cascade among planetesimals \citep{MustillWyatt2009, Sefilian2024}, and -- where planets and disks have inclination misalignments -- warp the vertical distribution of their dust \citep{Dawson_2011, Farhat2023, Sefilian_2025}. 
Recent high-resolution observations, in particular with the Atacama Large Millimeter/submillimeter Array (ALMA), have shown many examples of the expected emission morphologies in debris disk dust emission arising from such planet-disk interactions \citep[see e.g.,][]{MacGregor_2017, Marino_2018, Faramaz_2019, Matra_2019, Lovell_2021, MacGregor_2022, ImazBlanco_2023, Lovell_2025, Han_2026, Lovell_2026}.
Consequently, these findings justify continued studies of debris disk substructures, to search for planets via disks, and understand the broader evolution of exoplanetary systems. 

The focus of this work is the outer debris disk around $\epsilon$~Eridani (HD~22049), hereafter $\epsilon$~Eri, a K2V star in the constellation of Eridanus located at a distance of $d{=}3.22\,\rm pc$ \citep{2023A&A...674A...1G}. $\epsilon$~Eri has temperature $T_{\mathrm{eff}}{=}5052$\,K, stellar radius $R_\star{=}0.738\:R_\odot$, luminosity $L_\star{=}0.32\:L_\odot$ \citep{epsEri}, and mass $M_\star{=}0.82\:M_\odot$ \citep{Baines}, making it a very close analog to the Sun, at a younger age of 400--1200\,Myr \citep{Mamajek_2008,Sahlholdt_2018,Sanghi_2026}. 
$\epsilon$~Eri was one of the first stars around which a debris disk was detected \citep{Gillett1986}, and still remains the closest detected debris disk, which has been studied extensively at near-infrared, mid-infrared, far-infrared, and sub-/millimeter wavelengths \citep[see e.g.,][]{Aumann_1985, Greaves_1998, Greaves_2005, Backman_2009, Reidemeister_2010, Greaves_2014, MacGregor_2015, Chavez_Dagostino_2016, Booth_2017, Booth_2023, llopsayson2025, Wolff_2025}.
These observations altogether confirm the presence of a narrow (i.e., with a fractional width $\Delta R / R_\mathrm{mid} < 0.5$, corresponding to the lower quartile of belts in REASONS; \citealt{Matra_2025}), planetesimal belt centred at about 70\,au, from which smaller (${\sim}\mu$m-sized) dust is dragged inwards by Poynting-Robertson (P--R) and stellar wind (SW) drag.

$\epsilon$~Eri hosts one known planet, the Jupiter-analog $\epsilon$~Eridani~b first discovered  via radial velocity \citep{epsErib2000, epsEribKeck,epsEribGaia}. 
More recently, precise orbital measurements have refined the planetary parameters, with semi-major axis $a_\mathrm{pl}{=}3.53\pm0.04$\,au, inclination $i_\mathrm{pl}{=}40^{+6}_{-5}\deg$, a mass $m_\mathrm{pl}{=}1.00\pm0.10\, \mathrm{M_{Jup}}$, and eccentricity $e_\mathrm{pl} {=} 0.06 \pm 0.04$, consistent with a near-circular orbit \citep{thompson_2025}.
Additionally, multiple studies have predicted an additional planet, $\epsilon$~Eridani~c, based on the outer belt's structure. Namely, \citet{Booth_2023} interpreted low-surface-brightness azimuthal enhancements in the smoothed residuals as planetesimal clumping caused by a migrating planet trapped in a 2:1 MMR, predicting a planet with a mass of ${\sim}0.2\,M_{\rm Jup}$ at a semi-major axis of ${\sim}42$ au. On the other hand, \citet{Pearce_2022} inferred a potential planet sculpting the outer belt's inner edge, with a mass of ${\sim}0.19\,M_{\rm Jup}$ at ${\sim}52.7$ au. 
However, such planets have not yet been detected \citep[see, e.g.,][]{llopsayson2025, Wolff_2025}.
The non-detection is consistent with the sensitivity limits of the available observations: indeed, \cite{llopsayson2025} estimate that beyond ${\sim}16$ au, JWST Near Infrared Camera (NIRCam) data are sensitive to planets with masses of approximately Saturn's mass (${\sim}0.3\ M_{\rm Jup}$), above the mass inferred by both \cite{Booth_2023} and \cite{Pearce_2022}.

Given the similarities between the Sun and the star $\epsilon$~Eri, Jupiter and $\epsilon$~Eridani~b, and the Kuiper Belt and $\epsilon$~Eri's outer debris disk, the $\epsilon$~Eri system resembles a young Solar System analog, and may provide important clues as to the development of our own planetary system. In this work, we re-analyse ALMA Band~6 (1.29\,mm) observations of the $\epsilon$~Eri system.
We model the data with non-axisymmetric parametric disk models that separately incorporate forced and proper (or free) eccentricity terms, in order to test for a non-zero disk eccentricity. The computational methods, including model setup, and fitting procedures, are described in Section~\ref{sec:analysis}. Our results are presented in Section~\ref{sec:results} and discussed in Section~\ref{sec:discussion}, where the implications for potential planetary configurations are explored. We summarise and conclude our work in Section~\ref{sec:conc}.

\section{Data analysis \& Modelling} \label{sec:analysis}
\begin{figure}
\includegraphics[width=1.0\linewidth]{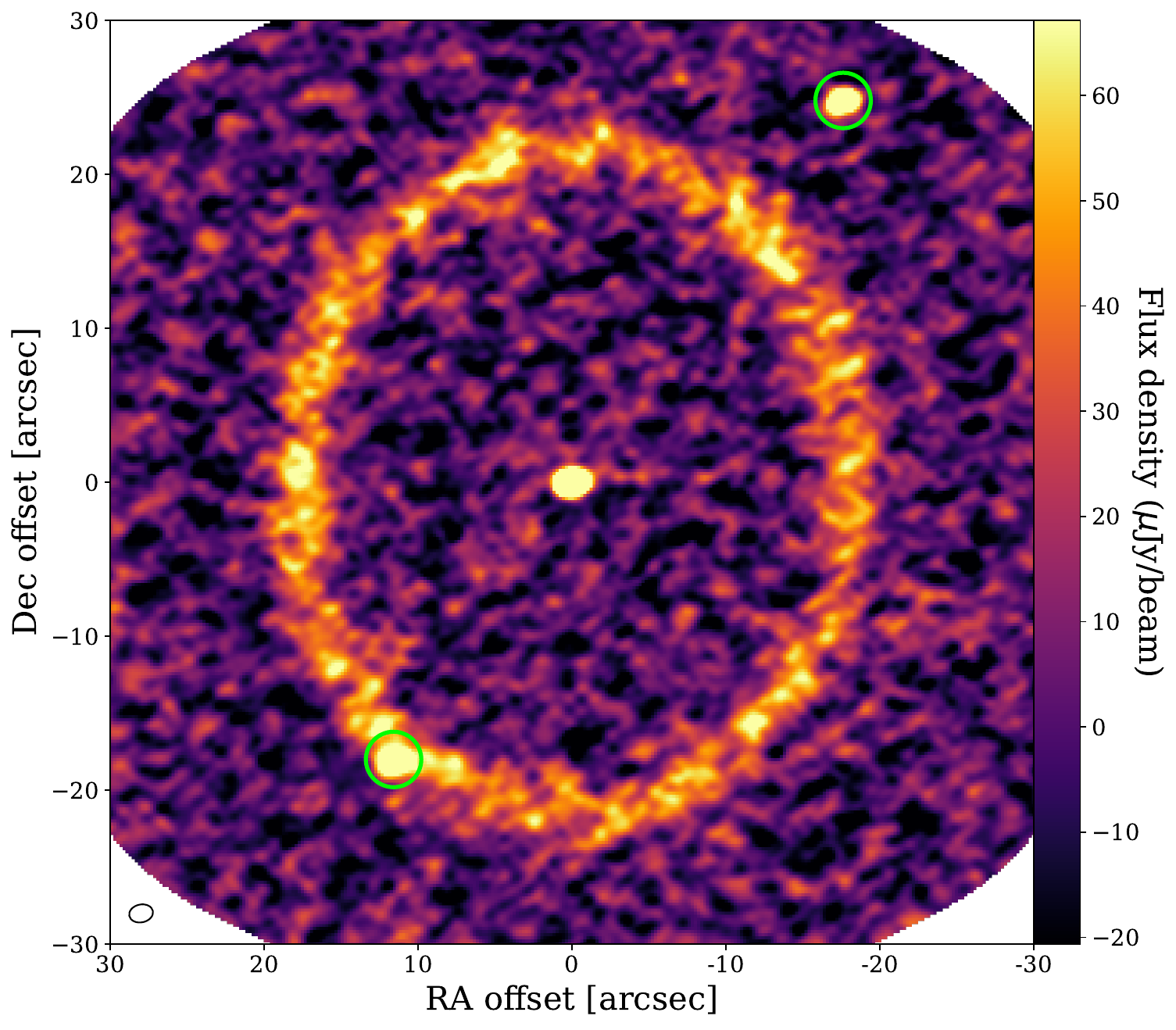}
\caption{ALMA 1.29\,mm image of the $\epsilon$~Eri system, adapted from \cite{Booth_2023}. A resolved debris disk and the star itself are visible. Two background sources  \citep[highlighted by green circles and identified as likely galaxies; see][]{Booth_2023} are detected to the north-west and south-east of the star, at offsets of approximately ($-18''$, $25''$) and ($12''$, $-18''$), respectively.
The size of the ALMA beam is shown in the bottom left. The image is oriented north-up and east-left.}
\label{fig:epseriIMAGE}
\end{figure}

In this work, we use the identical \texttt{CLEAN} image data products of $\epsilon$~Eri as presented in \cite{Booth_2023}. We show the image in Fig.~\ref{fig:epseriIMAGE}, and summarise key observation details below. $\epsilon$~Eri was observed by ALMA in Cycle~7 in project 2019.1.00696.S (PI: M. Booth), using Band~6 receivers (1.29 mm) between 2019 December 16 and 23. The total on-source integration time was 4.7 hours, split uniformly over a six-pointing hexagonal mosaic to cover the large angular extent of the disk, with mean precipitable water vapour levels of 1.1\,mm. The sensitivity of the primary beam is shown in the middle panel of Figure 1 presented in \cite{Booth_2023}.
The data were calibrated using the standard ALMA pipeline in \texttt{CASA} and imaged with \texttt{tclean} using natural weighting and a multiscale deconvolver, with the images achieving a peak (minimum) continuum sensitivity of $13.2\,\mu$Jy\,beam$^{-1}$. The final image has a pixel size of $0.2''$ and dimensions of $420{\times}420$ pixels. 
As seen in Fig.~\ref{fig:epseriIMAGE}, there are four main components: the outer belt, two background sources, which \cite{Booth_2023} identifies as galaxies, and the star itself, all of which are clearly resolved from each other. The background source to the south-east is co-located with the outer belt and the other to the north-west is separate from the disk's emission. 

\subsection{Model Setup} \label{subsec:model_setup}
To model the outer debris belt, we first define its surface density distribution. 
Following \cite{Lovell_2025}, we adopt a parametric description in which the disk is represented as a ring with a Gaussian radial surface density profile. 
Under this assumption, the surface density is given by:
\begin{equation}
    \Sigma(R, \phi; R_{\mathrm{mid}},\Delta R) = 
    \frac{1}{\sqrt{2\pi}\,\sigma j} 
    \exp\!\left[-\frac{(R - R_{\mathrm{mid}})^2}{2\sigma^2}\right],
    \label{eq:sigma_r}
\end{equation}
where $R$ is the radius from the star, $R_{\mathrm{mid}}$ is the mean radius of the outer debris disk, $\sigma$ is related to the full width at half-maximum (FWHM${=}\Delta R$) as $\Delta R {=} 2\sqrt{2\ln 2}\,\sigma$, and $j$ is the coordinate system Jacobian determinant as defined in \cite{Lynch2022}. 
For a circular, axisymmetric ring, one has $j {=} 1$, whereas for
an eccentric belt with a distance-independent eccentricity $e$, this becomes:
\begin{equation}
    j = \sqrt{1-e^2} .
\end{equation}
Generally, the total eccentricity $e$ of a particle's orbit may be decomposed into a forced component, $e_f$, and a proper component, $e_p$ \citep{1999ssd..book.....M}. These quantities are most naturally described as eccentricity vectors, with magnitudes $e_f$ and $e_p$ and orientations defined by their respective arguments of pericentre, $\omega_f$ and $\omega_p$. The forced eccentricity represents a coherent perturbation imposed on the disk by an external body, such as a planet, and results in particle orbits with a common argument of pericentre $\omega_f$. In contrast, the proper eccentricity describes the intrinsic eccentricity dispersion of individual particle orbits. In this case, particles possess a distribution of orbital orientations, such that the arguments of pericentre $\omega_p$ are uniformly distributed. Consequently, the forced eccentricity may lead to a global offset between the disk's geometric centre and the star, whereas the proper eccentricity broadens the radial structure of the disk without introducing a coherent offset \citep[e.g.,][]{Wyatt_1999, Kennedy_2020}. The total eccentricity $e$ is given by the magnitude of the vector sum of these two components:
\begin{equation}
    e = \sqrt{e_f^2+2e_fe_p\cos{(\omega_f-\omega_p)}+e_p^2}.
\end{equation}
Although perturbations may produce eccentricity distributions that vary with semi-major axis \citep[e.g.,][]{Wyatt_1999,Lynch2022, Sefilian2024}, we assume constant values of both $e_f$ and $e_p$ across the disk. This simplification is justified because the bulk eccentricity of $\epsilon$~Eri's outer belt is very low (see \S\ref{sec:results}), making the observations insensitive to radial variations in eccentricity.

To model the disk's vertical structure, following \citet{Lovell_2025}, we adopt a vertical Gaussian-distributed (linear) density profile:
\begin{equation}
    l_\rho(z)= \frac{1}{\sqrt{2\pi}H}\exp{\left(-\frac{z^2}{2H^2}\right)},
    \label{eq:l_z}
\end{equation}
where $H$ is the scale height, defined as $H{=}hR$, with $h$ being the vertical aspect ratio. 
This prescription produces a maximum density at the disk mid-plane ($z{=}0$) and a symmetric decrease in density above and below the mid-plane. Since the scale height increases linearly with radius, this prescription introduces an implicit radial dependence, such that the vertical density above/below the mid-plane decreases with increasing radius.
Combining the surface density and vertical linear density profiles of Equations (\ref{eq:sigma_r}) and (\ref{eq:l_z}) gives the full three-dimensional dust density distribution, $\rho {=} M_{\mathrm{dust}}\Sigma l_\rho$, with $M_{\mathrm{dust}}$ acting as a dust mass scaling factor that converts the model pixel intensities to physical mass units.

Using this parametrization, we define three debris disk models: 
(i) a circular model ($e_p{=}0$ and $e_f{=}0$), 
(ii) a model with forced eccentricity only ($e_p {=} 0$), and 
(iii) a model with proper eccentricity only ($e_f {=} 0$). 
The circular disk model serves primarily as a model validation case, as discussed in Appendix~\ref{sec:verification}. 
Whilst in principle it is possible to model both the forced and proper eccentricities together, given the constraints that we determine in Section~\ref{sec:results}, we do not consider such a scenario to avoid over-fitting.
\setcounter{footnote}{0}

The three-dimensional dust density model is provided as an input to the radiative transfer code \texttt{RADMC-3D}\footnote{The outer disk of $\epsilon$~Eri is safely within the optically thin regime at millimeter wavelengths due to its low surface density.} \citep{radMC}.
The dust population is modelled as a single grain species with a power-law size distribution, $dN/ds {=} s^{-3.5}$, consistent with a steady-state collisional cascade \citep{Dohnanyi}, as has been used extensively to model debris disks at sub-/millimeter wavelengths \citep[e.g.,][]{Marino_2015, Sibthorpe_2017}. 
The grain sizes are bounded between $0.9\,\mu$m and $1.0$~cm, a range chosen to span the sensitivity range of ALMA in Band~6.
The grain density is fixed to $2.7\,\mathrm{g\,cm^{-3}}$, consistent with compact silicate grains, as commonly adopted in the literature \citep[e.g.][]{Lovell_2021}. 
Vertically, the disk is assumed to be geometrically thin, with a fixed aspect ratio $h {=} H/R {=} 0.02$, consistent with the range of $h {\sim} 0.02 {-} 0.12 $ inferred for debris disks \citep{Hughes_2018, Zawadzki+2026}. 
The literature values for $\epsilon$~Eri's stellar mass, radius, and effective temperature (see \S\ref{sec:introduction}) are compared to Kurucz stellar spectra \citep{Kurucz1979} with the closest match providing an input to \texttt{RADMC-3D} to compute the dust temperature distribution via Monte Carlo radiative equilibrium calculations.
Although the dust temperature is modelled assuming this central radiation source, the star is strictly modelled as a point source with a flux assigned to it (which we denote as $F_{\nu,\epsilon\,\mathrm{Eri}}$).
Additionally, we model the background source to the south-east (Figure~\ref{fig:epseriIMAGE}) as an unresolved point source at a fixed sky position, determined from preliminary fitting, with its flux treated as a free parameter. 
We do not model the background source in the north-west as it is excluded by the mask we adopt (see \S\ref{sec:fitting}).
We simulate these models at the same wavelength as the ALMA Band~6 observations ($1.29$~mm). 
Each 3D model is then sky-projected and imaged in \texttt{RADMC-3D} by providing a source distance \citep[here we adopt $d{=}3.22$ pc;][]{2023A&A...674A...1G}, position angle (PA, measured anti-clockwise from north), inclination ($I$) and offset of the whole model in both RA ($\Delta$RA) and Dec ($\Delta$Dec). 
Finally, we convolve these images with the same beam as the ALMA data ($B_{\mathrm{maj}}{=} 1.57''$, $B_{\mathrm{min}} {=} 1.16''$, $B_{\mathrm{PA}} {=}-78.2$ deg) to produce synthetic ALMA images, which from hereon we refer to as `models'.

For ALMA data with resolution ${\gtrsim}0.15''$ the astrometric accuracy of ALMA data depends on the synthesised beam FWHM and the SNR of the data, as per Equation 10.7 of the ALMA handbook \citep{ALMAhandbook}.  
As highlighted in \cite{Booth_2023} where this data was first presented, since the central star has an SNR $> 20$, the mean absolute astrometric accuracy is $0.076''$.
Although absolute astrometric offsets would result in global offsets in the image, since in this work we model the disk and star relative to the disk coordinate centre, any absolute astrometric offsets in the data are accounted for by global model shifts via the $\Delta$RA and $\Delta$Dec parameters (which shift the star and disk model coordinates by the same amount).
This setup results in zero forced eccentricity and non-zero forced eccentricity scenarios having distinct values for $\Delta$RA and $\Delta$Dec, since by definition, the disk coordinate centre and loci of an eccentric ring are not co-located.
We note instead that relative astrometric errors, e.g., between the five execution blocks, and between the six-pointing mosaic, would influence the fitting, since these would result in emission smearing in the image, but that we consider these to be very low, given the point-source behaviour of the star, and uniformity of the outer belt in the image domain. 
The observations were conducted with a primary beam FWHM of $22''$ in six locations $18''$ from the phase centre \citep{Booth_2023}, and so the bright central star and a segment of the outer belt were always present within the primary beam main lobe FWHM in each pointing, resulting in the accurate reconstruction of the star and disk system, with no observable relative astrometric offsets.

\subsection{Model Fitting} \label{sec:fitting}
We adopt a fitting procedure consistent with \cite{Lovell_2025} and also fit in the image domain. 
We fit models to the ALMA data with Goodman \& Weare's Affine Invariant Markov chain Monte Carlo Ensemble sampler \citep{emcee}. 
The posterior distribution depends on the assumed prior distribution and a Gaussian likelihood function ($\propto \exp{(-\chi^2/2)}$).
We adopt the chi-squared ($\chi^2$) statistic as a measure of goodness-of-fit. 
The goodness-of-fit is evaluated by computing the $\chi^2$ between the beam-convolved model image and the primary-beam-corrected ALMA image, weighted by the local root mean square (RMS) noise. 
As image pixels contain correlated information, it is necessary to account for pixel-to-pixel correlations with a beam weighting factor, that effectively re-weights the noise by the number of pixels per beam. 
Accordingly, and as in \cite{Lovell_2025}, we adopt the following definition of $\chi^2$:
\begin{equation} \label{eq:chi2}
    \chi^2 = \sum_{x,y=1}^{N_{pix}} \frac{(I_{\mathrm{data}\:x,y}-I_{\mathrm{model}\: x,y})^2}{\sigma_{x,y}^2}\cdot \frac{\mathcal{M}_{x,y}}{\Omega} , 
\end{equation}
where $x,y$ are the pixel indices, $(I_{\mathrm {data}\ x,y}-I_{\mathrm{model}\ x,y})^2$ are the residuals squared, $\sigma_{x,y}$ is the error at the phase centre of observations scaled by the sensitivity map. We measured the image noise in the flat noise map over the largest possible emission-free elliptical region, obtaining $13.2\,\mu$Jy\,beam$^{-1}$, a value we adopt throughout this work. In Equation (\ref{eq:chi2}),
$\Omega$ is the beam area in pixels, defined as $\Omega {=} \pi B_{\mathrm{maj}}B_{\mathrm{min}}/(4\ln{2}) {\approx} 51.5\ \mathrm{pix}$ with $B_{\mathrm{maj}}$ and $B_{\mathrm{min}}$ being the beam FWHM along the major and minor axes, respectively. 
$\mathcal{M}_{x,y}$ is a binary mask, used to restrict the region over which the $\chi^2$ statistic is computed: this discards regions far from the disk that include only noise, thereby improving the precision of the calculation. 
Based on the properties of the data, we adopted a mask defined by the primary beam, requiring the primary beam level to be above 0.37; this excludes data with systematically poorer sensitivity, and ensures both that the background source in the north-west is excluded, and the star in the centre of the image is included.

We utilise priors to avoid walkers exploring non-physical regions of parameter space (e.g., negative fluxes and eccentricities), and to bound parameter exploration for disk properties that are already known with high precision (e.g., inclination). These priors entering the likelihood evaluation are uniform, as specified in the second column of Table~\ref{tab:mcmc_all_models}.
As recommended by {\tt emcee} documentation, we initialised parameters close to their final values after an initial model verification run, implemented to speed up the computation time without impacting the fitted results.

We assessed convergence using the integrated autocorrelation time ($\tau$), which provides an estimate of the number of steps required for successive samples to become effectively independent.
In our analysis, we used twice as many walkers as model parameters and a total number of steps exceeding $50{\times}\tau$, following the recommendations of \texttt{emcee}. 
We measured the auto-correlation time every 100 steps, to impose the $50{\times}\tau$ stopping condition without a priori knowing the total number of required steps.
The circular and forced-eccentric model converged after 8,500 steps and the proper-eccentric model after 13,500 steps. 
An initial portion of these walkers is discarded as `burn-in' to ensure that the walkers have had time to `forget' their initial position. 
In all cases, the first 50\% of the samples were conservatively discarded as burn-in. 
The typical auto-correlation time is ${\sim}200$, by discarding the first half of all chains, we are comfortably excluding the burn-in stage, whilst sufficiently sampling the posterior distribution for a good statistic.

\begin{figure*}
\includegraphics[width=\textwidth]{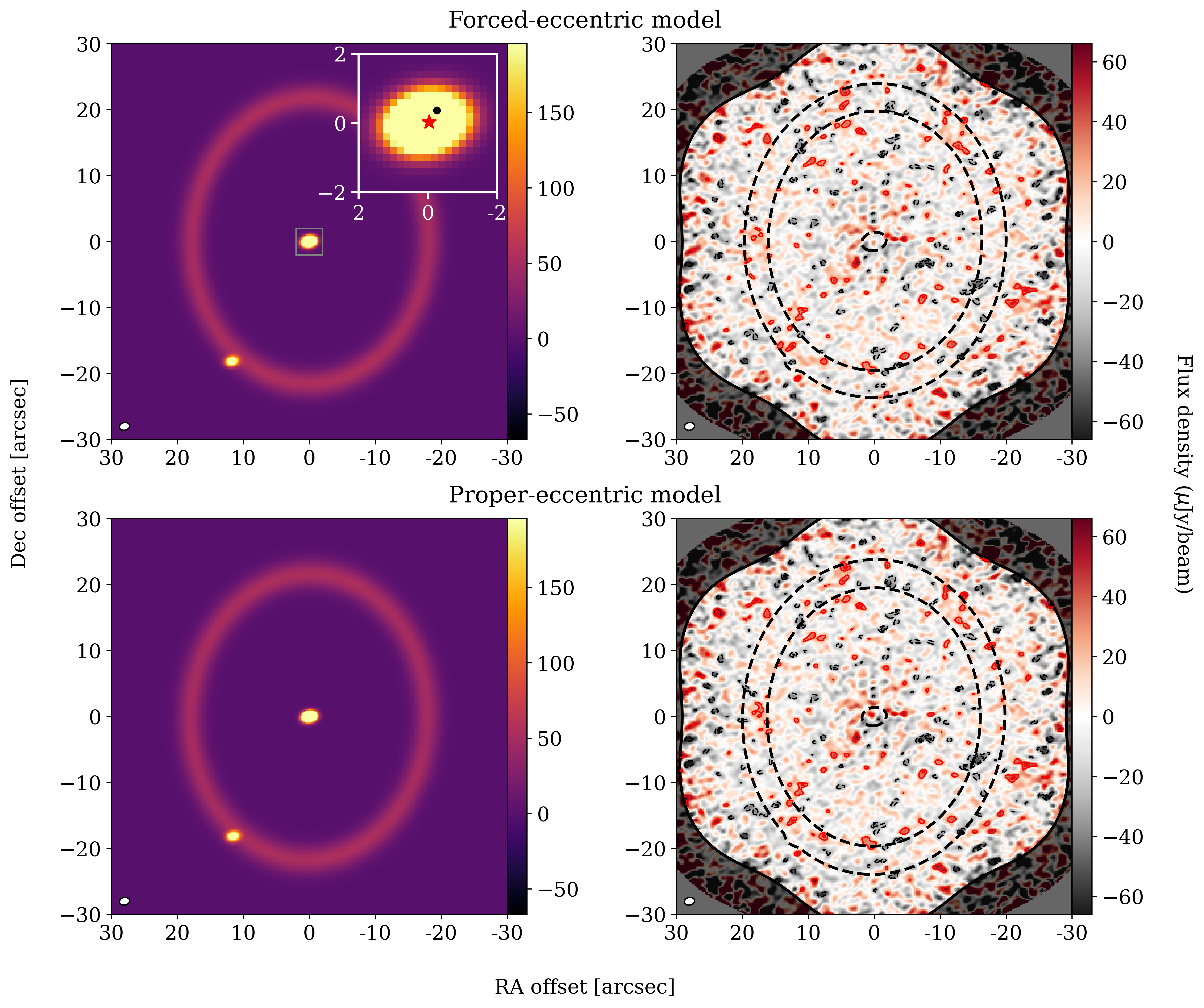}
\caption{\textit{Top left:} Best fit synthetic image for the Forced-eccentric model. The inset plot shows the location of the disk centre (black dot) and $\epsilon$~Eri (red star); the offset between these is driven by the best-fit value of $e_f$. \textit{Bottom left:} Best fit for the Proper-eccentric model. \textit{Right column:} Residuals, contours of $\pm2$ and $\pm3$ signal-to-noise ratio are shown. The dashed region indicates where the disk, star, and background source are located in the model, whereas the outer black line delineates the border of the fitting region. Size of the ALMA beam is shown in the bottom left of each panel. Results for the Circular disk model are presented in Appendix~\ref{sec:verification}.}
\label{fig:AllModels}
\end{figure*}

\begin{figure}
\includegraphics[width=1.0\linewidth]{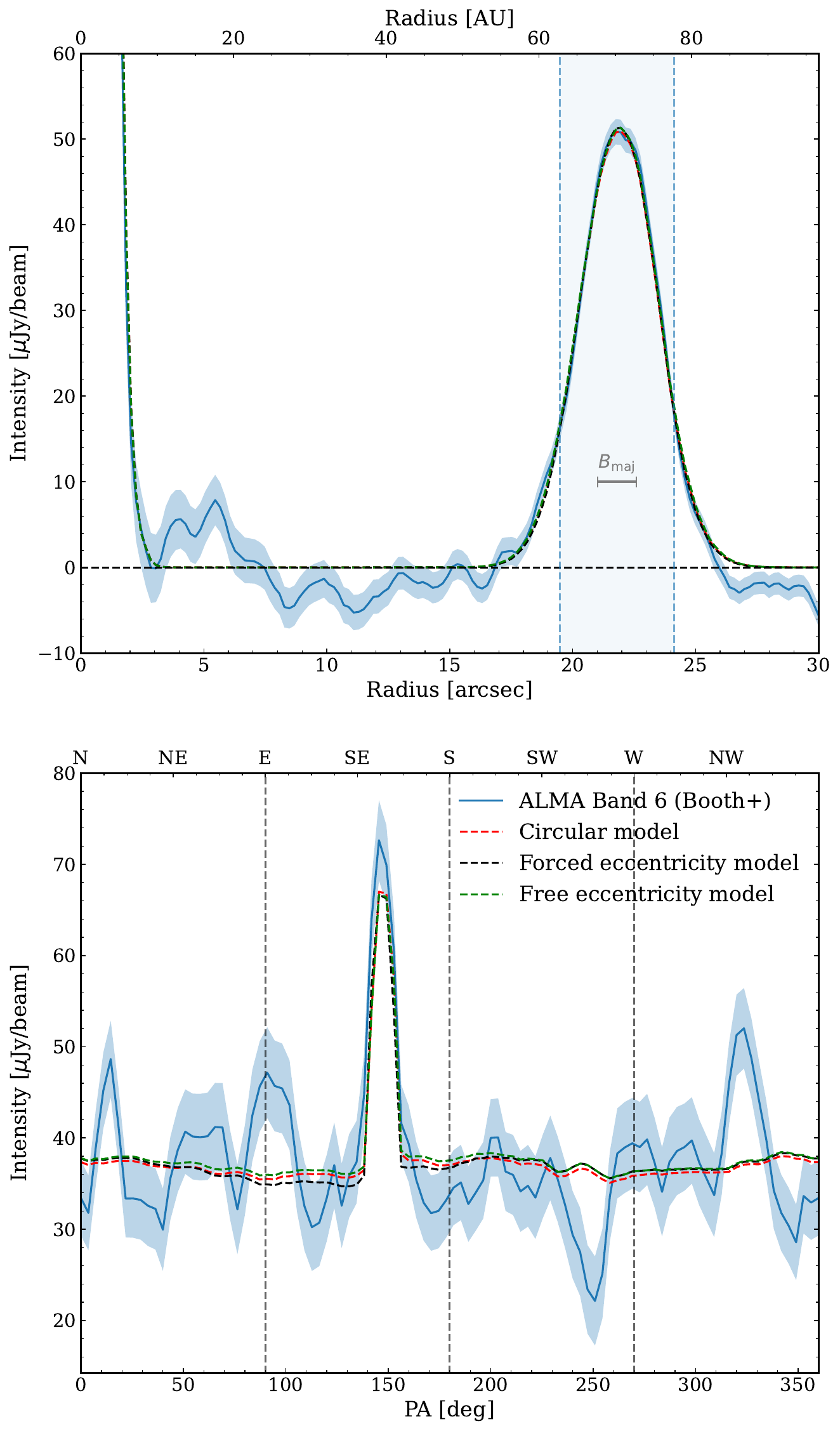}
\caption{Deprojected surface brightness profiles of the outer debris disk as a function of radial distance (top panel) and azimuthal angle (bottom panel). Profiles are shown for each of the three models considered in this work (using different coloured dashed curves), along with the data from \citet[][solid blue curve]{Booth_2023}.
For reference, the grey horizontal bar in the top panel labelled $B_{\mathrm{maj}}$ indicates the FWHM of the ALMA synthesized beam along its major axis.}
\label{fig:rad_az_model}
\end{figure}

\section{Results} \label{sec:results}
\subsection{Goodness of Fits}
We fitted the three specified models in Section~\ref{subsec:model_setup} to the ALMA data and in Table~\ref{tab:mcmc_all_models} we present the median and associated uncertainties of all model posterior parameters. 
Corner plots showing the (burn-in subtracted) posterior distributions for these can be found in Appendix~\ref{sec:corner}. 
All the best-fit models together with their residuals are shown in Fig.~\ref{fig:AllModels}.
The residuals are broadly consistent with noise in both the forced- and proper-eccentric cases, and the same is true for the symmetric case in Appendix~\ref{sec:verification}, with small differences in the precise locations of any ${\pm}2\sigma$ and ${\pm}3\sigma$ excesses.
We also plot the radial and azimuthal profiles of brightness for each model over the data in Fig.~\ref{fig:rad_az_model}. The profiles are constructed by deprojecting the image using the best-fit position angle and inclination, and then azimuthally averaging (for the radial profile) or radially averaging (for the azimuthal profile). We find that in the radial profile (top), the models lie within the $1\sigma$ uncertainties in the data. In the azimuthal profile (bottom), the models lie within the $3\sigma$ uncertainties in the data except for at PA${\sim} 320$ deg, where it is within $4\sigma$.
Overall, the 2-dimensional residuals and the 1-dimensional radial and azimuthal profiles demonstrate that in all cases, the models provide a very good description of the data.

\begin{deluxetable*}{rcccc}
\tablewidth{\textwidth}
\setlength{\tabcolsep}{23pt}
\tablecaption{Comparison of MCMC posteriors for the Circular, Forced-eccentric, and Proper-eccentric Gaussian ring models (\S\ref{subsec:model_setup}). Uncertainties represent the 16th and 84th percentile intervals. Dashes indicate parameters not applicable to a given model, and values in square brackets denote fixed parameters. The $\Delta\chi^2$ and $\Delta\mathrm{BIC}$ are calculated for the best-fit model relative to the Circular model presented in Appendix \ref{sec:verification}. \label{tab:mcmc_all_models}}
\tablehead{
\colhead{Parameter} & \colhead{Prior} & \colhead{Circular} & \colhead{Forced-eccentric} & \colhead{Proper-eccentric}
}
\startdata
$M_\mathrm{dust}$ [M$_\oplus$] & $(0,\ \infty)$ & $0.0487^{+0.0014}_{-0.0015}$ & $0.0486^{+0.0014}_{-0.0015}$ & $0.0487^{+0.0014}_{-0.0014}$ \\
$R_\mathrm{mid}$ [au] & $(15,\ 100)$ & $70.22^{+0.28}_{-0.29}$ & $70.14^{+0.28}_{-0.22}$ & $70.26^{+0.30}_{-0.30}$ \\
$\Delta R$ [au] & $(0,\ 50)$& $10.44^{+0.44}_{-0.47}$ & $10.32^{+0.44}_{-0.46}$ & $9.79^{+1.40}_{-0.73}$ \\
PA [deg] & $(-50,\ 50)$ & $-2.2^{+1.0}_{-1.0}$ & $-2.1^{+1.1}_{-1.1}$ & $-2.1^{+1.0}_{-1.1}$ \\
$I$ [deg] & $(5,\ 70)$ & $33.97^{+0.57}_{-0.57}$ & $33.77^{+0.59}_{-0.56}$ & $33.95^{+0.60}_{-0.59}$ \\
$F_{\nu,\epsilon\,\mathrm{Eri}}$ [mJy] & $(0,\ \infty)$ & $0.979^{+0.036}_{-0.037}$ & $0.976^{+0.035}_{-0.034}$ & $0.975^{+0.035}_{-0.035}$ \\
$\Delta$RA [arcsec] & $(-5,\ 5)$ & $-0.022^{+0.029}_{-0.028}$ & $0.029^{+0.032}_{-0.033}$ & $-0.023^{+0.029}_{-0.028}$ \\
$\Delta$Dec [arcsec] & $(-5,\ 5)$ & $0.053^{+0.024}_{-0.025}$ & $0.027^{+0.025}_{-0.024}$ & $0.053^{+0.025}_{-0.024}$ \\
$F_{\nu,\mathrm{bkgnd}}$ [mJy] & $(0,\ \infty)$ & $0.275^{+0.023}_{-0.023}$ & $0.282^{+0.022}_{-0.024}$ & $0.277^{+0.022}_{-0.022}$ \\
$x_\mathrm{bkgnd}$ [arcsec] & --- & $[11.52]$ & $[11.52]$ & $[11.52]$ \\
$y_\mathrm{bkgnd}$ [arcsec] & --- & $[-18.10]$ & $[-18.10]$ & $[-18.10]$ \\
$e_f$ & $(0,\ 1)$ & --- & $0.0162^{+0.0039}_{-0.0040}$ & --- \\
$\omega_f$ [deg] & $(0,\ 360)$ & --- & $125^{+12}_{-13}$ & --- \\
$e_p$ & $(0,\ 1)$ & --- & --- & $0.028^{+0.019}_{-0.022}$ \\
\hline
$\Delta\chi^2$ & --- & 0.0 & -18.8 & 0.2 \\
$\Delta\mathrm{BIC}$ & --- & 0.0 & -6.9 & 6.1 \\
\hline
\#Runs & --- & 8500 & 8500 & 13500 \\
Burn-in & --- & 4250 & 4250 & 6750 \\
\hline
\enddata
\tablecomments{The dust mass the model derives is approximately 10 times higher than expected based on the disk flux. This is likely a feature in how \texttt{RADMC-3D} treats masses provided by the surface density model. We investigated this and found that it had no impact on the morphological parameters, as $M_\mathrm{dust}$ simply represents a flux scaling factor. Moreover, both the data and models are comfortably within the optically thin regime, and the derived models reproduce the total disk flux very well.}
\end{deluxetable*}

\subsection{Parameter Posteriors}
The best-fit parameter posteriors in the Circular (validation), Forced-eccentric, and Proper-eccentric models
are self-consistent within $1\sigma$ 
for the fitted dust mass, disk radius, width, position angle, inclination, stellar flux, and background source flux, indicating good agreement among these bulk model components following convergence.
The posterior distributions differ primarily in the locations of the best-fit RA and Dec offsets and in the model-specific inputs, namely whether eccentricity was included. These differences are expected, as a forced eccentricity can introduce a global offset between the disk's geometric centre and the central star, resulting in different inferred offsets depending on the adopted model. We discuss these effects in the following subsections.
We also note here that a comparison between the Circular model parameters and those of \citet{Booth_2023} is provided in Appendix~\ref{sec:verification}. 

\subsubsection{Forced Eccentricity Model} \label{sec:forced_ecc}
For the Forced-eccentric model, the posterior distribution yields a nominal value of $e_f {=} 0.0162^{+0.0039}_{-0.0040}$, a result at the level of $4\sigma$ significance; see Fig.~\ref{fig:histogram_eccentricity}. The corresponding displacement of the inferred disk centre relative to the stellar position is shown in the inset of top left panel of Fig.~\ref{fig:AllModels}.
The posterior distribution of the argument of forced eccentricity ($\omega_f$) exhibits a preferred orientation, $\omega_f{=} 125^{+12}_{-13}\ \mathrm{deg}$ (measured anti-clockwise from north), which points approximately in the direction of the background source located within the disk region. This alignment could indicate that the background source is biasing the inferred eccentricity. 
To test whether this is the case, we performed an additional fit of the Forced-eccentric model in which we masked out the background source in the image, excluding this small region from the fitting (a mask ${1.5}\times$ the beam dimensions about the peak location of the background source was used).
We present the posterior distributions of this fit in Fig.~\ref{fig:forced_corner_masked}, and note here that the resulting values of $e_f$ and $\omega_f$ remain consistent with those obtained when we included the background source.
Overall, this suggests that the background source alignment with $\omega_f$ is coincidental.
Whilst the posteriors suggest that the model has tentatively derived a forced eccentricity in $\epsilon$~Eri's outer belt, i.e., at the level of $4\sigma$, we are cautious about this finding and interpret this as an upper-limit, given that this has not passed a ${>}5\sigma$ significance threshold.
We therefore adopt a conservative upper limit of $e_f < 2.6\%$, defined as the value below which 99.7\% ($3\sigma$) of the posterior samples lie (Fig.~\ref{fig:histogram_eccentricity}). 

We note as an aside that given the disk orientation (inclination and position angle) and the uncertainties in $\omega_f$, the probability that a randomly located background source within the disk would be found in the preferred (tentative) pericentre direction is $6.5\%$. 
Whilst this value is low and therefore unlikely, clearly we cannot reject the alignment as happening purely by chance.

Given the proper motion of $\epsilon$~Eri \citep[$\mu_{\alpha*} {=} -974.758 \pm 0.160 \ \mathrm{mas\,yr^{-1}}$, $\mu_{\delta} {=} +20.876 \pm 0.120 \ \mathrm{mas\,yr^{-1}}$;][]{2023A&A...674A...1G} and the time since these observations were obtained, this south-east background source should now be separated from the disk emission.
As such, follow-up ALMA observations would no longer have to model this source/exclude it from fitting.
We finally note that deeper and/or higher-resolution ALMA observations could improve upon the constraint derived by this model, and thus potentially enable a more direct measurement of $e_{\rm f}$, rather than an upper limit.

\subsubsection{Proper Eccentricity Model} \label{sec:proper_ecc}
\begin{figure}
\includegraphics[width=\linewidth]{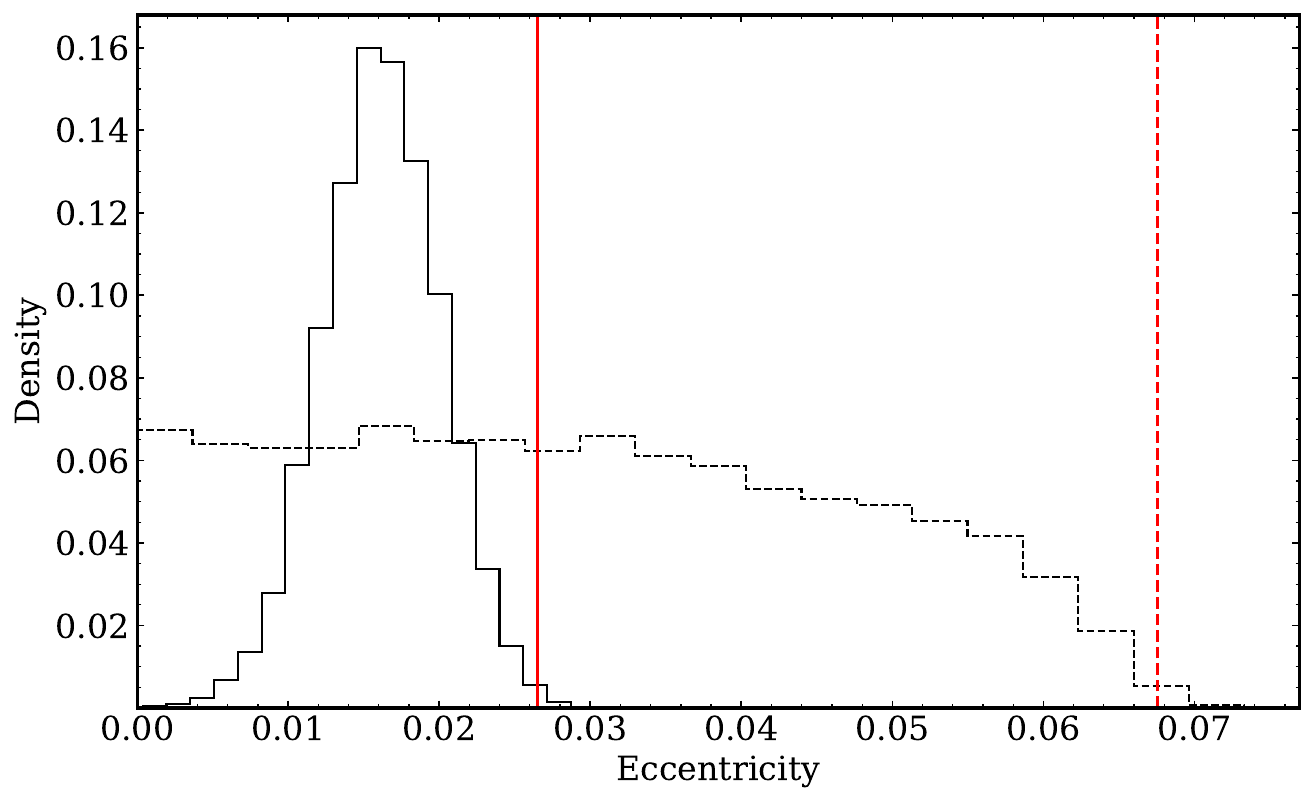} 
\caption{Normalized histograms of the forced (solid line) and proper (dashed line) eccentricities. The red vertical lines indicate the corresponding 99.7\% ($3\sigma$) upper bounds, showing that  99.7\% of samples are within ${\sim}0.026$ for the Forced-eccentric model and ${\sim}0.068$ for the Proper-eccentric model.}
\label{fig:histogram_eccentricity}
\end{figure}

The posterior distribution of $e_p$ in the Proper-eccentric model is more straightforwardly interpreted as an upper-limit, and yields a value of $e_p{=} 0.028^{+0.019}_{-0.022}$, see Fig.~\ref{fig:histogram_eccentricity}. 
These posteriors are clearly non-Gaussian, due to the degeneracy of $e_p$ and $\Delta R$ \citep[a known effect in debris disks that are not well resolved across their radii; see, e.g.,][]{MacGregor_2017, MacGregor_2022, Lovell_2023} and consequently, these posterior mean and variance values are not meaningful summary statistics.
We adopt an upper limit of $e_p < 6.8\%$ based on the 99.7th percentile value in the $e_p$ posteriors. Whilst this is a less stringent constraint than that placed on the forced eccentricity, it still provides a robust limit on the proper eccentricity of the disk, given the data available.

\subsection{Model Comparison}
We now compare the three different models considered. We emphasize, however, that these models are not necessarily competing descriptions of the same physical scenario, as they constrain different aspects of the belt's eccentricity structure. 
We begin by noting that whilst the $\Delta \chi^2$ values demonstrate that the Forced-eccentric model provides the best-fit to the data (with $\Delta \chi^2{=}{-18.8}$; Table \ref{tab:mcmc_all_models}), this alone does not establish that it is the preferred model, given the differing complexities of the models considered.
As such, to determine whether one model provides a statistically superior description of the data, we make use of the Bayesian information criterion (BIC), which offers a more appropriate comparison than $\Delta \chi^2$ since this applies a penalty to a given model's complexity, i.e. the number of free parameters. 
Namely, BIC is defined as:
\begin{equation} \label{eq:BIC}
    \mathrm{BIC} = \chi^2 +k \ln{(N_{\mathrm{dat}})},
\end{equation}
where $k$ is the number of free parameters and $N_{\mathrm{dat}}$ is the number of independent data points, and thus penalizes models with more free parameters \citep{BIC}. 
In our analysis, the $\chi^2$ value is taken from the best-fit result from the converged {\tt emcee} chains, $N_{\mathrm{dat}}$ corresponds approximately to the number of independent resolution elements (beams) within the fitting region, which is $N_{\mathrm{dat}} {\simeq} 379$ for the adopted fitting region.\footnote{A more stringent mask, encompassing fewer noise-dominated regions, would reduce $N_{\mathrm{dat}}$, thereby lowering the $\Delta \rm BIC$ values for the Forced-eccentric model and increasing it for the Proper-eccentric model.} The three models have $k{=}9$ (Circular), $k{=}10$ (Proper-eccentric) or $k{=}11$ (Forced-eccentric) free parameters.
We compared models based on the difference in their BIC values, as:
\begin{equation} \label{eq:dBIC}
\Delta \mathrm{BIC} = \Delta \chi^2 + \ln(N_{\mathrm{dat}})(k_i - k_j),
\end{equation}
where $i$ and $j$ denote two competing models. 
Overall a lower BIC value indicates a preferred model, with $\Delta \mathrm{BIC}$ providing a quantitative measure of the strength of the preference.

Despite the BIC penalization of models with more parameters, in Table~\ref{tab:mcmc_all_models} we can see that there is still a slight preference for the Forced-eccentric model, as it has the smallest $\Delta \rm BIC$ value of ${-6.9}$, in comparison to the Proper-eccentric model which has a positive $\Delta \rm BIC$ value of ${6.1}$, where a value of $\Delta \rm BIC{=}0.0$ is measured with respect to the Circular model.
In general, $\Delta \rm BIC$ values of $\geq$ 10 are often considered `very strong' evidence favouring the lower $\mathrm{BIC}$ model.
Whilst this means that this quantitative metric does not provide strong evidence to accept/reject the hypothesis that the Forced-eccentric or Proper-eccentric models are statistically preferred to the Circular model, we note that the $\Delta \rm BIC$ between the Forced-eccentric and Proper-eccentric models is $13.0$, suggesting that the Forced-eccentric model is statistically favoured over the Proper-eccentric model.

To summarize, all three models provide good fits to the ALMA data; although a marginal (${\sim}4\sigma$) forced eccentricity signal is suggested, it is not significant enough to claim a detection and is instead treated as an upper limit of $e_f < 2.6\%$, while the proper eccentricity is also constrained as an upper limit of $e_p < 6.8\%$, with model comparison (via BIC) showing only weak preference for the Forced-eccentric model over the circular case, but a clear preference over the Proper-eccentric model.

\section{Discussion} \label{sec:discussion}

\subsection{A Near-circular Debris Ring}

The inferred low forced and proper eccentricities for $\epsilon$~Eri's outer debris disk imply that the disk has little to no discernible coherent eccentric structure and is consistent with a dynamically `cold' planetesimal population, analogous to that of the cold classical Kuiper belt \citep[CCKB; e.g.,][]{Morbi2020}. The CCKB is commonly characterised by a low-inclination population with $i\leq4^\circ$ \citep[e.g.,][]{DelsantiJewitt2006}; adopting the approximate equipartition between eccentricity and inclination excitation of \cite{IdaMakino1992}, $\sqrt{\langle e^2\rangle}{=}2\sqrt{\langle i^2\rangle}$, this corresponds to $\sqrt{\langle e^2\rangle}\lesssim0.14$, well above the eccentricity upper limits derived for $\epsilon$~Eri's outer belt. 
This suggests that the disk has not been strongly forced into an eccentric configuration by e.g. planet-induced secular perturbations \citep{Wyatt_1999}, nor strongly excited through self-stirring \citep{IdaMakino1992}.
We note in passing, however, that low forced eccentricity need not uniquely imply weak secular perturbations. 
In particular, secular interactions between planets and less massive debris disks may damp forced eccentricities of planetesimals while leaving free eccentricities unaffected through a process known as secular resonant friction \citep{Sefilian2023}.
Interestingly, the inferred upper limit on the forced eccentricity is lower than that on the free eccentricity (Fig. \ref{fig:histogram_eccentricity}), a result that may support this interpretation. However, because both $e_p$ and $e_f$ are constrained only by upper limits, the present data do not establish their relative magnitudes. Higher-resolution observations will therefore be required to investigate this possibility further.

Overall, the disk appears markedly more quiescent than many well-studied eccentric debris disks with reported forced eccentricities of $e_{f} {\sim} 0.12{-}0.21$ that can be visible by eye \citep[such as Fomalhaut, HD~202628, and HD~53143, see e.g.][]{MacGregor_2017, Faramaz_2019, MacGregor_2022, Lovell_2026}, and thus suggests that the $\epsilon$~Eri planetary system has undergone a different evolutionary history.
The modelling yields a conservative 99.7th percentile upper limit of $e_{f} < 0.026$, which, to the best of our knowledge, represents the lowest upper-limit constraint on the eccentricity of a debris disk reported to date from millimeter observations. 
Typical millimeter observational upper limits for debris rings are of order $e_{\rm f} {\sim} 0.05{-}0.10$ \citep[such as for $\eta$~Corvi and q$^1$~Eri, see][]{Marino+2017, Lovell_2021}. 
The near-percent level precision therefore suggests these millimeter data are more than a factor of two more sensitive to underlying belt eccentricities than previous (low) upper limits.

The upper limits derived here also place $\epsilon$~Eri in a distinct dynamical regime from the small population of narrow, eccentric debris rings such as Fomalhaut, HD~202628 or HD~53143 \citep{Kennedy_2020}. For an eccentric ring, the radial excursion associated with the forced eccentricity is expected to introduce a characteristic fractional width of order $\Delta R / R_\mathrm{mid} {\sim} 2e_f$. Systems such as Fomalhaut and HD~202628 are notable because they appear significantly narrower than this expectation, implying that additional dynamical mechanisms (or formation pathways) may be required to maintain their narrow morphologies. In contrast, our constraints indicate that for $\epsilon$~Eri, $\Delta R/R_\mathrm{mid}{\approx}15\%$, and thus $2e_f \lesssim \Delta R/R_\mathrm{mid}$, such that the disk is not anomalously narrow for its allowed eccentricity. In this respect, $\epsilon$ Eri appears more consistent with systems such as $\eta$~Corvi, HD~181327, or q$^1$~Eri, rather than the unusually confined eccentric rings, as discussed by \cite{Kennedy_2020}.

It is also instructive to compare the results derived from these ALMA data with those obtained at other wavelengths for $\epsilon$~Eri. 
The James Webb Space Telescope Mid-Infrared Instrument (JWST/MIRI) observations provide a substantially weaker constraint on the disk eccentricity of $0.09^{+0.04}_{-0.05}$ \citep{Wolff_2025}. 
Whilst we note that the underlying models used to derive eccentricities employed in this work and in \citet{Wolff_2025} differ, the weaker constraint (albeit one that is consistent with the ALMA results to within $2\sigma$) likely reflects the more diffuse nature of mid-infrared thermal emission, which traces smaller grains with broader spatial distributions and no clear outer edge, which are likely spiralling inwards via stellar wind drag. Furthermore, the MIRI data are impacted by diffraction spikes and plausibly imperfect stellar source subtraction owing to saturation of the star. Instead, the ALMA emission traces the well-defined parent planetesimal belt and directly detects the precise location of the star, making it well-suited for constraining subtle geometric offsets.
In contrast, based on a north-south flux asymmetry observed in Herschel PACS (Photodetector Array Camera and Spectrometer) and SPIRE (Spectral and Photometric Imaging Receiver) images, \citet{Greaves_2014} reported $\epsilon$~Eri's eccentricity as ${>}0.03$, inferring this to result from the pericentre glow effect, and ${<}0.1$ based on the absence of a significant offset between the stellar location and the disk centre.
The results of this study contradict the lower-limit reported by \citet{Greaves_2014}, given the maximal eccentricity in the posterior distribution of 0.026, and suggests that instead of resulting from pericentre glow, the lower-resolution Herschel images may have suffered from background confusion.

\subsection{The $\epsilon$~Eri Exoplanetary System: A single giant planet?} \label{sec:planetB}
We now consider whether $\epsilon$~Eri~b alone  can explain the measured eccentricity of the outer belt.
An important consideration is whether secular perturbations from $\epsilon$~Eri~b have had sufficient time to influence the disk over the lifetime of the system.
We estimate the secular timescale $t_{\rm sec}$ at the midpoint of the debris ring $R_{\rm mid}$, using the free apsidal precession rate of a massless test particle due to an interior planet of mass $m$ and semi-major axis $a$. According to the linear Laplace--Lagrange theory, this rate is given by \citet{1999ssd..book.....M}:
\begin{equation} \label{eq:Ap} 
A_p = \frac{1}{4} \sqrt{\frac{G M_\star}{R_{\rm mid}^3}} \frac{m}{M_\star} \frac{a}{R_{\rm mid}} b_{3/2}^{(1)}(\alpha)
\end{equation} 
where $b_{3/2}^{(1)}(\alpha)$ is a Laplace coefficient with $\alpha {\equiv} a/R_\mathrm{mid}$, 
\begin{equation} \label{eq:laplace} 
b_s^{(j)}(\alpha) = \frac{1}{\pi} \int_0^{2\pi} \frac{\cos(j\psi)\,d\psi} {\left(1 - 2\alpha\cos\psi + \alpha^2\right)^s}. 
\end{equation} 
The corresponding secular timescale is $t_\mathrm{sec} {=} 2\pi/A_p$. 
Using the measured mass and semi-major axis of $\epsilon$~Eri~b \citep{thompson_2025}, we find that $t_\mathrm{sec}{\approx} 292~{\rm Myr}$, which is shorter than the minimum estimated age of the system \citep[$400~{\rm Myr}$;][]{Mamajek_2008} and substantially less than the mean estimated age that we adopt later ($800~{\rm Myr}$). 
Therefore, the disk has had sufficient time to dynamically respond to perturbations from $\epsilon$~Eri~b. 

We can then evaluate whether the measured disk eccentricity can be reproduced by the forced eccentricity induced by $\epsilon$~Eri~b. 
The forced eccentricity at $R_{\rm mid}$ due to an interior planetary perturber is given by \citep{1999ssd..book.....M}:
\begin{equation} \label{eq:ef}
    e_f =\frac{b_{3/2}^{(2)}(\alpha)}{b_{3/2}^{(1)}(\alpha)} e_{\rm pl}\approx \frac{5}{4}\frac{a}{R_{\rm mid}}e_{\rm pl},
\end{equation}
where $e_{\rm pl}$ is the orbital eccentricity of $\epsilon$~Eri~b, and the approximate form is the $\alpha {\to} 0$ limit.
Adopting the maximum eccentricity for $\epsilon$~Eri~b of $10\%$ \citep{thompson_2025}, we obtain a forced eccentricity of $e_f {\simeq} 0.63\%$. This value is consistent with our upper limit of $e_f < 2.6\%$ and lies within the distribution of fitted $e_f$ values inferred from the eccentric disk model presented in Section~\ref{sec:forced_ecc}. Specifically, an $e_f$ of $0.63\%$ or less is recovered in approximately $0.89\%$ of samples from the posterior distribution. Therefore, provided that $\epsilon$~Eri~b has a non-zero eccentricity, the known planet alone is capable of reproducing the inferred distribution of fitted $e_f$ values. In this limiting case, there is no strict dynamical requirement for an additional outer planet to force the inferred low eccentricity into the outer debris disk.

\subsection{The $\epsilon$~Eri Exoplanetary System: An additional outer planet?} \label{sec:planet_c_diff_constraints}
Since the eccentricity of $\epsilon$~Eri~b is only weakly constrained observationally and may be consistent with a circular orbit, we now explore whether an additional planet located exterior to $\epsilon$~Eri~b but interior to the outer belt could instead be responsible for exciting the disk eccentricity.
As discussed in Section~\ref{sec:introduction}, such an outer planet, $\epsilon$~Eri~c, has previously been proposed based on both the reported clumpy structure of the disk \citep{Booth_2023} and the location of its inner edge \citep{Pearce_2022}. These studies inferred broadly similar planet masses of ${\sim}0.2\,M_{\rm Jup}$, but different semi-major axes of $42$ and $52.7$ au, respectively.
Given the structure of the outer belt, in this section we place constraints on the plausible mass ($m$) and semi-major axis ($a$) of any outer planet (beyond $\epsilon$~Eri~b). These constraints are based on the ALMA data, together with recently published JWST/MIRI data \citep{Wolff_2025}, Spitzer Infrared Array Camera (IRAC) \citep{Janson2008,Janson2015} and JWST/NIRCam \citep{llopsayson2025} direct imaging, and additional dynamical arguments outlined below.

We begin by noting that the existence of the outer belt itself excludes the presence of any sufficiently massive planet within the radial extent of the belt. We define the disk boundaries as $R_{\rm in}{=}R_{\rm mid}{-}3\sigma$ (${\approx} 57$ au) and $R_{\rm out}{=}R_{\rm mid}{+}3\sigma$ (${\approx}83$ au), corresponding to the orange shaded region shown in Figure~\ref{fig:par_space}. Any planet located within this region would be expected to strongly perturb, scatter, or deplete the planetesimal belt on timescales short compared to the system age.

\subsubsection{Secular perturbations} \label{subsec:sec_planet_C}

In this subsection, we consider a hypothetical Planet~c as the sole secular perturber of the outer belt, neglecting the dynamical influence of $\epsilon$~Eri~b.
We first note that secular perturbations from massive planets imprint measurable forced eccentricities on debris disks over timescales much longer than individual orbital periods. The lack of a measurable forced eccentricity thus implies that any perturbing body must either have a relatively low eccentricity, if located near the disk, or be well-separated from the planetesimal belt.
We estimate the region of parameter space over which a planet can secularly perturb the disk by requiring that the secular perturbation timescale at the disk radius, $R_\mathrm{mid}$, be shorter than or comparable to the age of the system, i.e., $t_{\rm sec} \lesssim t_{\rm age}$.
This condition defines the region in which a putative planet would have had sufficient time to dynamically influence the outer belt over the system lifetime. 
Using the secular precession timescale defined in Section~\ref{sec:planetB}, setting $t_{\rm sec}{=}t_{\rm age}$, and solving for the planet mass gives:
\begin{equation}
    m = \frac{8\pi}{t_\mathrm{age}}\sqrt{\frac{M_\star R_\mathrm{mid}^3}{G}}\frac{R_\mathrm{mid}}{a}\left[b_{3/2}^{(1)}(\alpha)\right]^{-1}.
\end{equation}

\begin{figure*}
\includegraphics[width=1.0\linewidth]{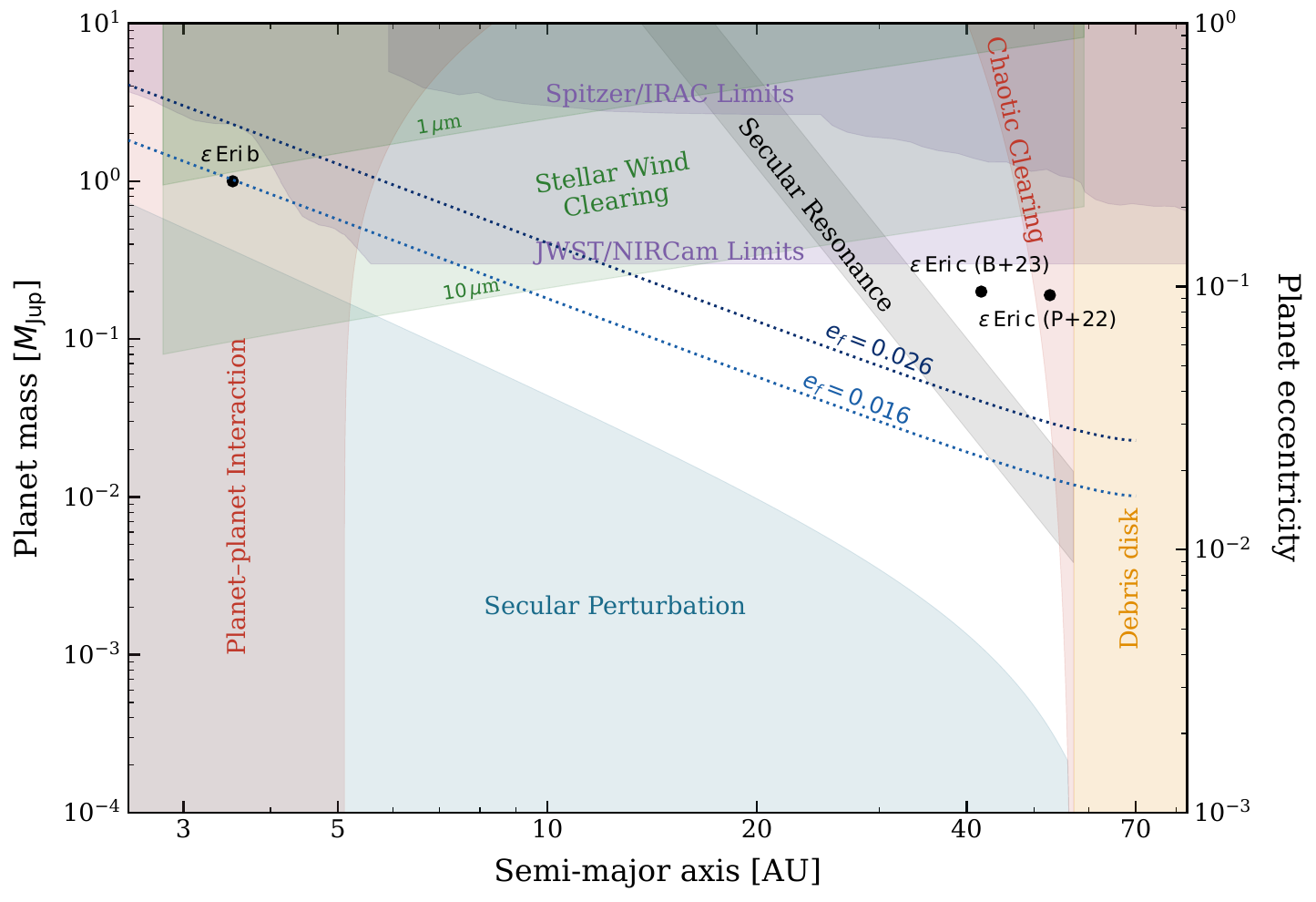}
\caption{Mass versus semi-major axis exclusion diagram for a putative planet in the outer regions of the $\epsilon$~Eri system. The white area indicates combinations of planet mass and semi-major axis consistent with imaging non-detections and different dynamical arguments (for details, see \S\ref{sec:planet_c_diff_constraints}).
The stellar wind clearing region is shown for two different dust particle sizes, as indicated in the plot. 
The two blue dotted lines show, as a function of planetary semi-major axis, the planetary eccentricity (read off the secondary y-axis, which applies only to these blue curves) required to reproduce a given forced eccentricity in the disk (labelled above each line) via secular forcing as described by Equation (\ref{eq:ef})}.
\label{fig:par_space}
\end{figure*}

For simplicity and illustrative purposes, we adopt $t_\mathrm{age}{=}800$ Myr, corresponding to the mean of the range given in Section~\ref{sec:introduction}.
The condition $t_\mathrm{sec}{=}t_\mathrm{age}$, corresponds to the limiting case where a planet has had just enough time to secularly perturb the disk over the system lifetime. Planets with longer secular timescales ($t_\mathrm{sec}>t_\mathrm{age}$) would therefore lie within the disallowed blue region labelled `Secular Perturbation' in Fig.~\ref{fig:par_space}.
If the age reported by \cite{Sanghi_2026} were instead adopted ($t_\mathrm{age}{=}1.1$ Gyr), the inferred mass limits would decrease by approximately 27\%, given the inverse dependence of the mass limit on system age.

Given that $t_{\rm sec} \lesssim t_{\rm age}$ for part of the $(a,m)$ parameter space, we next determine the eccentricity required for Planet~c to reproduce representative values of the fitted disk forced eccentricity, namely $e_f {=} 0.016$ and $0.026$, i.e., the model best-fit value and upper-limit value from Section~\ref{sec:forced_ecc}. 
To do so, we make use of the exact form of Equation~(\ref{eq:ef}) and solve for the planet's eccentricity.
Because the forced eccentricity is independent of planet mass, these constraints are represented as a function of planet semi-major axis alone and are shown on the secondary y-axis of Fig.~\ref{fig:par_space}.

Before moving on, we note that several caveats apply to these calculations. 
First, both $t_{\rm sec}$ and $e_f$ are computed assuming a single perturbing planet, neglecting the presence of $\epsilon$~Eri~b, which may also contribute to the secular forcing of the disk.
Second, the calculations assume a massless disk. Including a non-zero disk mass would modify the inferred constraints 
such that the planet masses derived here should be regarded as lower limits at a given semi-major axis \citep{Sefilian2024}.

\subsubsection{Secular resonances} \label{sec:secular}
If two planets are present, their mutual secular interactions, together with their combined perturbations on debris disk particles, can give rise to secular apsidal resonances. This occurs when the free apsidal precession rate of the planetesimals matches one of the eigenfrequencies of the two-planet system \citep{1999ssd..book.....M}. At such locations, the forced eccentricity can grow substantially (i.e., $e_f {\to} 1$), leading to the sculpting of a gap within the disk \citep{Yelverton2018, Sefilian2021}. The absence of any obvious gap structure in the outer belt therefore provides an additional constraint on the properties of a putative outer companion. A caveat here is that, given the resolution and sensitivity of the data, a gap narrower than 5 au with 50\% clearing could be present, but would not be resolvable given the existing data. We confirmed this by measuring the residual radial profiles obtained by subtracting a disk model with a gap from the data. Although \citet{Yelverton2018} demonstrate that secular resonance-sculpted gaps are cleared by 50\% \citep[see also][]{Sefilian2021, Sefilian2023}, it is non-trivial to predict the widths of these gaps\footnote{This is in contrast to resonance widths, which can be calculated analytically \citep[e.g.][]{Yelverton2018}. As discussed in \citet{Sefilian2021}, however, these do not necessarily correspond to observable gap widths, which depend non-trivially on the underlying planetesimal distribution, including their eccentricities and apsidal angles.}, as doing so requires e.g. N-body simulations, which are beyond the scope of this work.

\citet{Yelverton2018} derive the location, time-scale, and width of these resonances using classical Laplace--Lagrange secular theory, under two main assumptions:
(i) the disk is represented by massless test particles, so the planetesimals feel the planets' gravity but exert none themselves; and (ii) the theory is only valid to second order in eccentricities, so the resulting forced eccentricities can grow without bound near a resonance, which is unphysical. This means the theory is reliable for identifying where a resonance-driven depletion should appear, but $N$-body simulations are needed to establish what the resulting gap actually looks like in detail. Relatedly, including a non-zero disk mass modifies the secular precession frequencies and thus the resonance locations \citep{Sefilian2021,Sefilian2023}, potentially altering the inferred planetary constraints. None of these idealizations is expected to change the qualitative constraint we use here, but they do mean that our excluded planetary mass at fixed semi-major axis should be read as an approximate, rather than precise, boundary.

Following \citet{Yelverton2018}, and adopting their Equation~(24), which is derived assuming $a_\mathrm{pl} \ll a \ll r_\mathrm{gap}$, we estimate the planetary mass required for a secular resonance to occur at a gap location $r_\mathrm{gap}$ within the disk.
Solving for the outer planet mass $m$ gives:
\begin{equation}
m = m_\mathrm{pl} a^{-11/2} r_\mathrm{gap}^{7/2} a_\mathrm{pl}^{2},
\end{equation}
where $m_\mathrm{pl}$ and $a_\mathrm{pl}$ denote the mass and semi-major axis, respectively, of $\epsilon$~Eri~b. We therefore consider resonance locations spanning the full extent of the observed outer belt, $R_{\rm in}\leq r_{\rm gap}\leq R_{\rm out}$. Since $m\propto r_{\rm gap}^{7/2}$, the lower bound $r_{\rm gap}=R_{\rm in}$ corresponds to the lowest planetary mass capable of placing the resonance within the belt at a given semi-major axis $a$, while $r_{\rm gap}=R_{\rm out}$ corresponds to the highest such mass. Placing a resonance at $r_\mathrm{gap}$ is only a meaningful exclusion if the resonance also has time to act within the age of the system. We therefore additionally checked, using Equation~(25) of \citet{Yelverton2018}, that the secular time-scale at the resonance is shorter than the system age for every $(a,m)$ combination excluded on resonance-location grounds alone; this was satisfied in all cases. Consequently, all combinations of $(a,m)$ that would place the resonance within the observed outer belt can in principle be excluded, since no corresponding gap is observed. In practice, this allows one to rule out planets capable of generating secular resonances between the inner and outer edges of the disk, thereby providing an additional constraint corresponding to the grey region labelled `Secular Resonance' in Fig.~\ref{fig:par_space}.

The discussion thus far has assumed non-migrating planets. Migrating planets are treated by \citet{Yelverton2018} only as a separate, exploratory scenario. They find that planetary migration can cause a sweeping secular resonance to widen the resulting depletion relative to the static case, at the cost of increased disk asymmetry. However, this result also highlights an important constraint: a planet cannot have migrated (either outward or inward) through the secular resonance band at any point in its history without potentially disturbing the outer debris disk.

In closing, secular resonances should be taken into account when considering a two-planet scenario. If the orbital parameters of $\epsilon$ Eri b had been different, the resulting secular architecture could have placed a resonance such that $\epsilon$ Eri c -- as hypothesized by \citet{Booth_2023} and \citet{Pearce_2022} -- would have been excluded.

\subsubsection{Planet-planet interactions}
We utilize the mutual Hill radius around $\epsilon$~Eri~b using the revised orbital parameters \citep{thompson_2025} to place an additional dynamical stability constraint on the presence of a second planet. Two planets on nearby orbits are expected to become dynamically unstable if their orbital separation is smaller than a few mutual Hill radii. Following \citet{KOKUBO1998}, we require the orbital spacing to exceed 5 mutual Hill radii to ensure stability. 
Solving for mass we obtain:
\begin{equation}
m = 3M_\star
\left(
\frac{2}{5}
\frac{\Delta a}{a+a_\mathrm{pl}}
\right)^3
-
m_\mathrm{pl},
\end{equation}
where $\Delta a {=} |a - a_\mathrm{pl}|$ is the separation from $\epsilon$~Eri~b. Planets with a mass higher than this limit are expected to be dynamically unstable on relatively short timescales and are therefore excluded from the allowed parameter space shown in Fig.~\ref{fig:par_space} by the red `Planet-planet Interaction' region.

\subsubsection{Chaotic clearing} \label{sec:chaotic}
We may also use the observed location of the debris belt's inner edge to constrain planets through chaotic clearing. The width of a chaotic zone is defined according to Equation~(1) of \cite{Quillen2006}, which follows the resonance overlap criterion originally derived by \cite{Wisdom_1980}, as:
\begin{equation} \label{eq:chaotic}
    \frac{R_\mathrm{in}-a}{a} = 1.5\left(\frac{m}{M_\star+m}\right)^{2/7}.
\end{equation}
This expression defines the red region in Fig.~\ref{fig:par_space}, labelled `Chaotic Clearing'. Along this boundary, the outer edge of a planet's chaotic zone coincides with the observed inner edge of the debris belt. Planets lying on this boundary could therefore be responsible for sculpting the belt's inner edge. In contrast, planets located closer to the outer belt would possess wider chaotic zones that extend into the observed belt, clearing material beyond $R_\mathrm{in}$ and shifting the inner edge outward. Such configurations are therefore inconsistent with the observed disk morphology.

\subsubsection{Stellar wind clearing}
The JWST/MIRI detection of the warmer inner debris disk \citep{Wolff_2025} places restrictions on the parameter space in which exoplanets can exist due to interactions with dust transported inwards via stellar wind drag, though we note these as being highly uncertain because the grain population responsible for the observed mid-infrared emission is not known a priori. 
From the 25.5\,$\mu$m radial surface brightness profile presented by \cite{Wolff_2025}, the inner belt appears smooth between ${\sim} 2.8$ and ${\sim} 59$ au. The inner bound is defined by the smallest radius probed by the observations, while the outer bound corresponds to the radius at which the signal-to-noise ratio decreases to unity. The constraints on planetary architectures are primarily informed by the region where the surface brightness profile has sufficient sensitivity to reveal potential substructures. In particular, between approximately 5 and 30\,au, where the SNR ranges from 1000 down to 10 respectively, and this constraint dominates over those from other observational limits, the radial surface brightness profile has no significant depressions or discontinuities that could be produced by planetary clearing.
The absence of resolved emission substructures (e.g., rings, gaps, or shoulders) across this region implies that any planet residing between these radii must clear material on a timescale shorter than the replenishment timescale of dust dragged inwards from the edge of any nearby, outer substructure, i.e., preventing the formation of any persistent, observable substructures, on resolution scales comparable to those of JWST/MIRI ($0.803''$ or $2.59\,\mathrm{au}$).
We therefore approximate the outer edge of a potentially unresolved gap as $a_\mathrm{out} {\approx} a + \delta a/2$, where $\delta a {=} 2.59$ au. 

The lack of any observed gap therefore places an upper limit on the width of any planet-induced cleared zone. 
Using the clearing timescale as stated in Equation~(3) of \cite{Shannon_2016} and equating it to the combined P--R and SW timescale ($t_\mathrm{drag}$) from Equation~(B14) of \cite{Wolff_2025}, we solve for the minimum planet mass:
\begin{equation} \label{eq:SW_mass}
    \frac{m}{M_\oplus} = \left[\frac{t_\mathrm{drag}}{\alpha}\left(\frac{a}{1 \ \mathrm{au}}\right)^{-\beta}
   \sqrt{{\frac{M_{\odot}}{M_\star}}
   }
    \right]^{\frac{1}{\gamma}},
\end{equation}
where $\alpha {=} 4\times10^6\ \mathrm{yr}$, $\beta {=} 1.6$, and $\gamma {=} -0.94$ are empirically determined coefficients by \cite{Shannon_2016}. The drag timescale is given by \citep{Wolff_2025}:
\begin{equation} \label{eq:t_drag}
t_\mathrm{drag}=
    \frac{a_\mathrm{out}^2 - a^2}{4GM_\star(\beta_\mathrm{rad}/c+\beta_\mathrm{sw}/v_\mathrm{sw})},
\end{equation}
which corresponds to the time required for dust grains to migrate from $a_\mathrm{out}$ to $a$ under the combined effects of P--R and SW drag. $\beta_\mathrm{rad}$ and $\beta_\mathrm{sw}$ are both grain-size dependent and are taken from the left panel of Figure 10 of \cite{Wolff_2025} for grain sizes ($a_\mathrm{grain}$) of 1 and 10 $\mu$m, corresponding to the median and largest grains considered in that work. The minimum grain size shown in that figure (0.1 $\mu$m) yields substantially weaker constraints on planet masses (i.e., ${\gtrsim}10\,M_{\rm Jup}$) and therefore does not meaningfully restrict the planetary parameter space, given the upper-limits implied by Spitzer/IRAC (see \S\ref{sec:disc_dirIm}). We adopt $v_\mathrm{sw} {=} 400$\,km\,s$^{-1}$, the stellar wind velocity used by \citet{Wolff_2025}. Using Equations (\ref{eq:SW_mass}) and (\ref{eq:t_drag}), we exclude the green `Stellar Wind Clearing' regions in Fig.~\ref{fig:par_space}.

\subsubsection{Direct imaging} \label{sec:disc_dirIm}
Direct imaging limits are taken from the mass-sensitivity curve of \citet{Janson2008,Janson2015} for an assumed system age of 800 Myr. Since these limits are reported as a function of projected separation, we convert them to physical separations by dividing by the cosine of the inclination of the best-fit Forced-eccentric model. This de-projection assumes that any putative planet is co-planar with the outer belt and follows a circular orbit. It also corresponds to the most conservative geometry, in which the planet is located along the projected minor axis of the disk (i.e., perpendicular to the line of nodes), where the difference between projected and true separation is maximized. The de-projected sensitivity curve was then plotted onto the exclusion diagram shown in Fig.~\ref{fig:par_space}. The resulting constraint is shown as the purple region labelled `Spitzer/IRAC Limits'.

We additionally include constraints from the JWST/NIRCam F444W observations of \citet{llopsayson2025}. We digitized the published $5\sigma$ mass-sensitivity curve corresponding to the 150\,K clear atmosphere model of \citet{Linder2019} from their Figure 5, and extrapolated this outwards. While the published mass-sensitivity curve extends only to $2''$, where it already reaches sensitivities below a Saturn mass, \citet{llopsayson2025} further state that their observations preclude Saturn-mass planets at separations greater than $16\,$au. As the contrast sensitivity flattens, or continues to improve with increasing separation, we adopt a conservative approximation by imposing a constant Saturn-mass detection limit beyond ${\sim}1.7''$ to avoid adding a discontinuous jump in the JWST/NIRCam limit that we present. We note that, if the system age of $1.1\pm0.1$ Gyr as proposed by \cite{Sanghi_2026} holds, the planet may have cooled to temperatures at which water clouds become important. Such clouds can suppress the ${\sim}4\, \mu$m emission and thereby weaken the resulting mass constraints. Nevertheless, even under this conservative treatment, the JWST/NIRCam limits appear to be significantly more constraining than those from the earlier Spitzer/IRAC observations and stellar wind clearing from micron-sized dust beyond ${\sim}4\,$au, whilst still remaining insensitive at the $5\sigma$ level to a planet with $\epsilon$~Eri~b's mass and semi-major axis. The resulting constraint is shown in Fig.~\ref{fig:par_space} as the purple region labelled `JWST/NIRCam Limits'.

\subsubsection{The exoplanet exclusion zones of $\epsilon$~Eri}
By combining these multiple constraints, we make reference to the exoplanet exclusion diagram on Fig.~\ref{fig:par_space}.
Even considering these six distinct constraints, the allowed parameter space remains quite large, ranging at least from ${\sim} 2\times10^{-4}$ to $0.3\,M_\mathrm{Jup}$ in planet mass and ${\sim} 6{-}60$ au in semi-major axis. The upper mass bound is primarily set by the JWST/NIRCam direct imaging limits, while the lower mass bound is determined by the secular perturbation criterion.
The stellar wind clearing constraint depends sensitively on the assumed grain size, $a_\mathrm{grain}$. By first assuming the most stringent of these stellar wind clearing limits ($a_{\rm grain}{=}10\, \mu$m), the exclusion diagram implies the possible presence of planets with masses and semi-major axes consistent with Saturn, Uranus and Neptune within the system, with relatively low, maximal eccentricities of ${\approx} 0.1$. However, we note that this limiting case would also be inconsistent with the proposed properties of $\epsilon$~Eri~b, indicating that adopting a grain size of  $10\, \mu$m in this stellar wind model is implausible.
If instead $a_{\rm grain}{=}1\,\mu$m was adopted, the stellar wind clearing constraint is substantially relaxed. In either case, however, the allowed parameter space is now bounded above by the JWST/NIRCam direct imaging limits, which exclude planets more massive than approximately a Saturn mass at wide separations. Consequently, once the direct imaging constraint is included, the choice of $a_\mathrm{grain}$ no longer has a significant effect on the overall excluded region.
Nonetheless, since lower-mass planets induce relatively weak observational signatures and typically fall below the sensitivity limitations of existing instruments, those are currently too challenging to detect.

Further progress will require observations that target both the disk morphology and the planet detection limits. For example, deeper, higher-resolution ALMA observations would refine measurements of the disk eccentricity, either by tightening the inferred forced eccentricity upper-limit, or perhaps confirming the tentatively derived value of $1.62\%$. Additionally, such observations would place tighter constraints on the presence of any cleared gaps narrower than 5 au, which could then be used to determine, or more strictly constrain the properties of the putative $\epsilon$ Eri c.
Deeper JWST/NIRCam observations would likewise be valuable, as the current NIRCam mass limits are close to the masses proposed for $\epsilon$ Eri c. Complementary JWST/NIRCam and JWST/MIRI point-source sensitivity curves (extending beyond $2''$) would help to constrain wide-separation planetary masses excluded in this system, while future ELT-class facilities could further probe the remaining parameter space through high-contrast imaging and improved long-period companion searches.

Finally, we note that the constraints on the planetary architecture presented in this work are based on a simplified analytic approach that provides reasonable first-order limits on the masses and orbital locations of potential companions. A more comprehensive dynamical analysis, for example using N-body simulations to explore the long-term evolution of the system, would provide a more complete assessment of the permitted planetary architectures and represents a promising avenue for future work.

\section{Conclusions} \label{sec:conc}
We present eccentric models of $\epsilon$~Eridani's outer debris belt at millimeter wavelengths by re-analysing ALMA Band 6 data presented in \citet{Booth_2023} and fitting parametric debris disk models in accordance with \citet{Lovell_2025} and \citet{Lynch2022}. A circular Gaussian ring model was first employed as a validation exercise, demonstrating that clean image fitting can robustly reproduce the disk parameters previously derived from dirty image fitting. The close agreement with the results of \cite{Booth_2023} confirms that the adopted modelling framework is reliable. Building upon this baseline, we explored more complex disk architectures motivated by secular planet--disk interactions, allowing for both forced and proper eccentricity components. Our conclusions are as follows:

\begin{enumerate}[label=(\roman*)]

    \item Accounting for the increased model complexity via the Bayesian information criterion, we find a mild preference for the Forced-eccentric model over the Circular model ($\Delta\mathrm{BIC}{=}6.9$) and a strong preference over the Proper-eccentric model ($\Delta\mathrm{BIC}{=}13.0$). The eccentric models provide physical upper-limits on the forced and proper eccentricity components, important for assessing the dynamical state of the outer belt.
    
    \item From the Forced-eccentric model, we derive a 99.7th percentile upper limit on the disk eccentricity of $e_f < 2.6\%$ with a posterior mean value of $e_f {=} 0.0162^{+0.0039}_{-0.0040}$. The Proper-eccentric model yields a similarly low constraint, with a 99.7th percentile upper bound of $e_p < 6.8\%$ and a posterior mean value of $e_p {=} 0.028^{+0.019}_{-0.022}$. Together, these results indicate that the outer debris belt of $\epsilon$~Eridani is nearly circular and dynamically cold. 

    \item Combining these constraints with planet--disk interaction theory, stellar wind drag models, JWST/MIRI observations of the warm component of the debris disk \citep{Wolff_2025}, and JWST/NIRCam F444W constraints \citep{llopsayson2025}, we place limits on potential additional planets in the system. The observations remain consistent with a system containing only the known planet, $\epsilon$~Eri~b, and do not require the presence of an additional planet. Nevertheless, the permitted parameter space for an outer planet remains relatively broad, ranging at least from $2\times10^{-4}$ to $0.3\, M_\mathrm{Jup}$ in mass and $6$--$60$ au in semi-major axis, and includes planets with masses and orbits comparable to exo-Neptunes or exo-Saturns.
    
\end{enumerate}

\begin{acknowledgments}
OU acknowledges funding from Trinity College Dublin.
JBL acknowledges the Smithsonian Institution for funding via a CfA \textit{J.C. Ryan} Fellowship.
AAS is supported by the Heising-Simons Foundation through a 51 Pegasi b Fellowship.
We thank the anonymous referee for their careful reading of the manuscript, and constructive comments.
We thank Markus Janson for helpful discussions and assistance with the interpretation of the Spitzer/IRAC results.
We thank Anna Fehr for useful discussions on an earlier version of Figure~5.
This paper makes use of ALMA data ADS/JAO.ALMA\#2019.1.00696.S. ALMA is a partnership of ESO (representing its member states), NSF (USA) and NINS (Japan), together with NRC (Canada), NSC and ASIAA (Taiwan),and KASI (Republic of Korea), in cooperation with the Republic of Chile. The Joint ALMA Observatory is operated by ESO, AUI/NRAO and NAOJ.
The National Radio Astronomy Observatory  is a facility of the U.S. National
Science Foundation operated under cooperative agreement by Associated Universities, Inc.
\end{acknowledgments}

\begin{contribution}
\textbf{Oto Ulrich}: Formal analysis, Investigation, Visualization, Writing - Original Draft. 
\textbf{Joshua B. Lovell}: Conceptualization, Data Curation, Methodology, Project administration, Software, Supervision, Validation, Writing - Original Draft. 
\textbf{David J. Wilner}: Conceptualization, Funding acquisition, Project administration, Supervision, Writing - Review \& Editing.
\textbf{Antranik A. Sefilian}: Methodology, Validation, Writing - Review \& Editing.
\textbf{Mark Booth}: Data curation, Resources, Writing - Review \& Editing.
\end{contribution}

\facilities{ALMA}
\software{astropy \citep{astropy:2022},
CARTA \citep{angus_comrie_2025_17050846},
corner \citep{Foreman-Mackey2016},
emcee \citep{emcee}, 
matplotlib \citep{Hunter:2007},
numpy \citep{harris2020array},
RADMC-3D \citep{radMC}}.

\appendix

\section{Model validation}\label{sec:verification}

\begin{figure*}
\includegraphics[width=\textwidth]{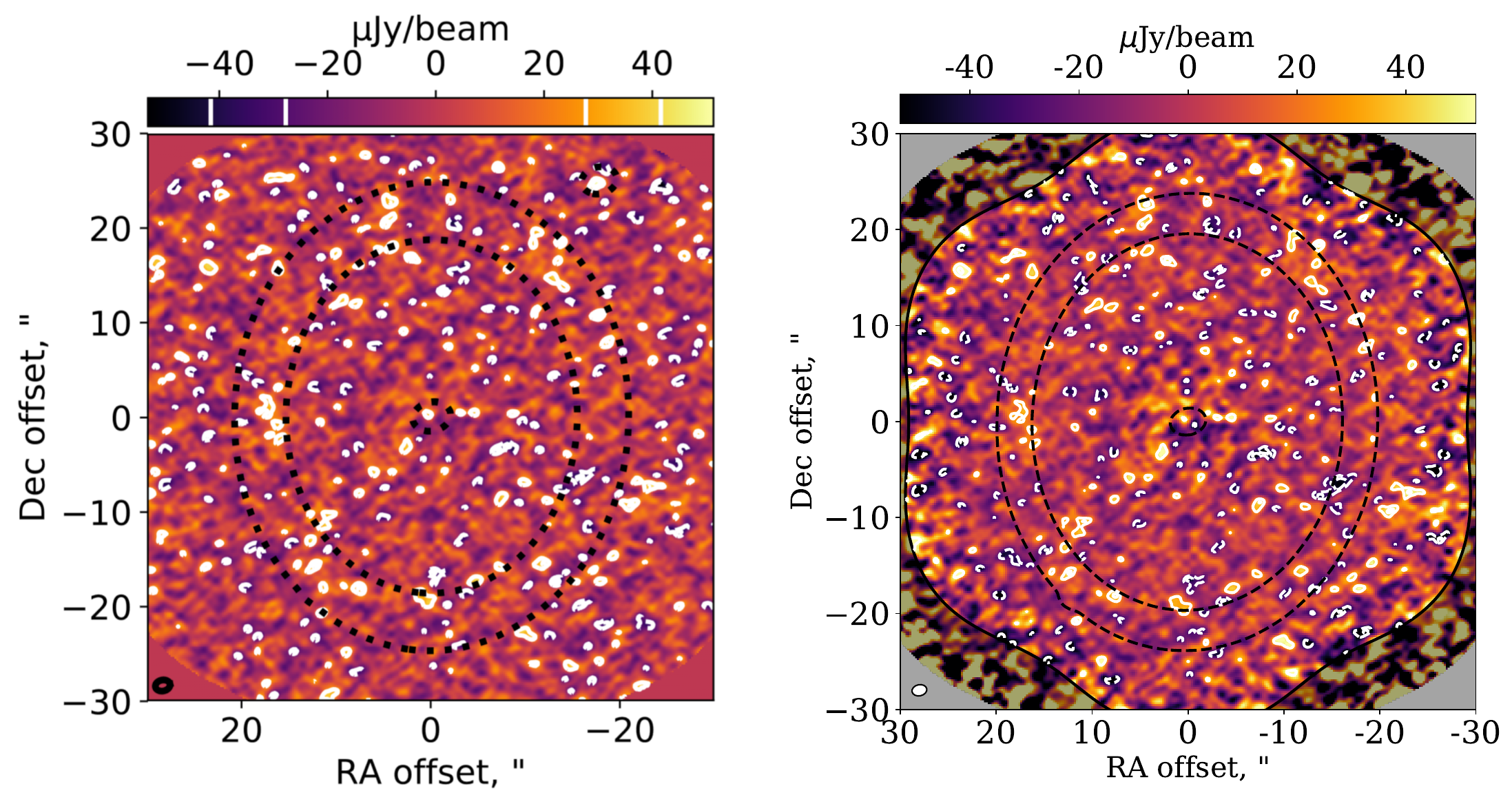}
\caption{\textit{Left:} Residuals for the circular model fitted by \citet{Booth_2023} to the dirty image. \textit{Right:} Residuals for the Circular model fitted in this work to the primary beam corrected clean image. Contours of $\pm2$ and $\pm3$ signal-to-noise ratio are shown, with consistent residual structures recovered by both fitting approaches. In each panel, the size of the ALMA beam is shown in the bottom left.}
\label{fig:residuals_comparison}
\end{figure*}

We verify the modelling and fitting procedure outlined in Section~\ref{sec:analysis} by comparing the results we obtain for the Circular model with those presented in \cite{Booth_2023}. Its role is to demonstrate that the modelling framework can independently reproduce previously published results and recover well-established disk parameters. A comparison between the residuals from subtracting the best-fit model (derived in this work) from the data and those produced by \cite{Booth_2023} is presented in Fig.~\ref{fig:residuals_comparison}.
By comparing the left and right panels of Fig.~\ref{fig:residuals_comparison}, one can see that the two modelling approaches yield near-identical residual maps. Quantitatively, this is further supported by the fact that the best-fit parameters reported in Table \ref{tab:mcmc_all_models} herein and in Table 3 of \cite{Booth_2023} agree within 1$\sigma$, indicating that the two approaches are consistent with one another.

\section{Posterior Distributions} \label{sec:corner}
\begin{figure*}
\includegraphics[width=\textwidth]{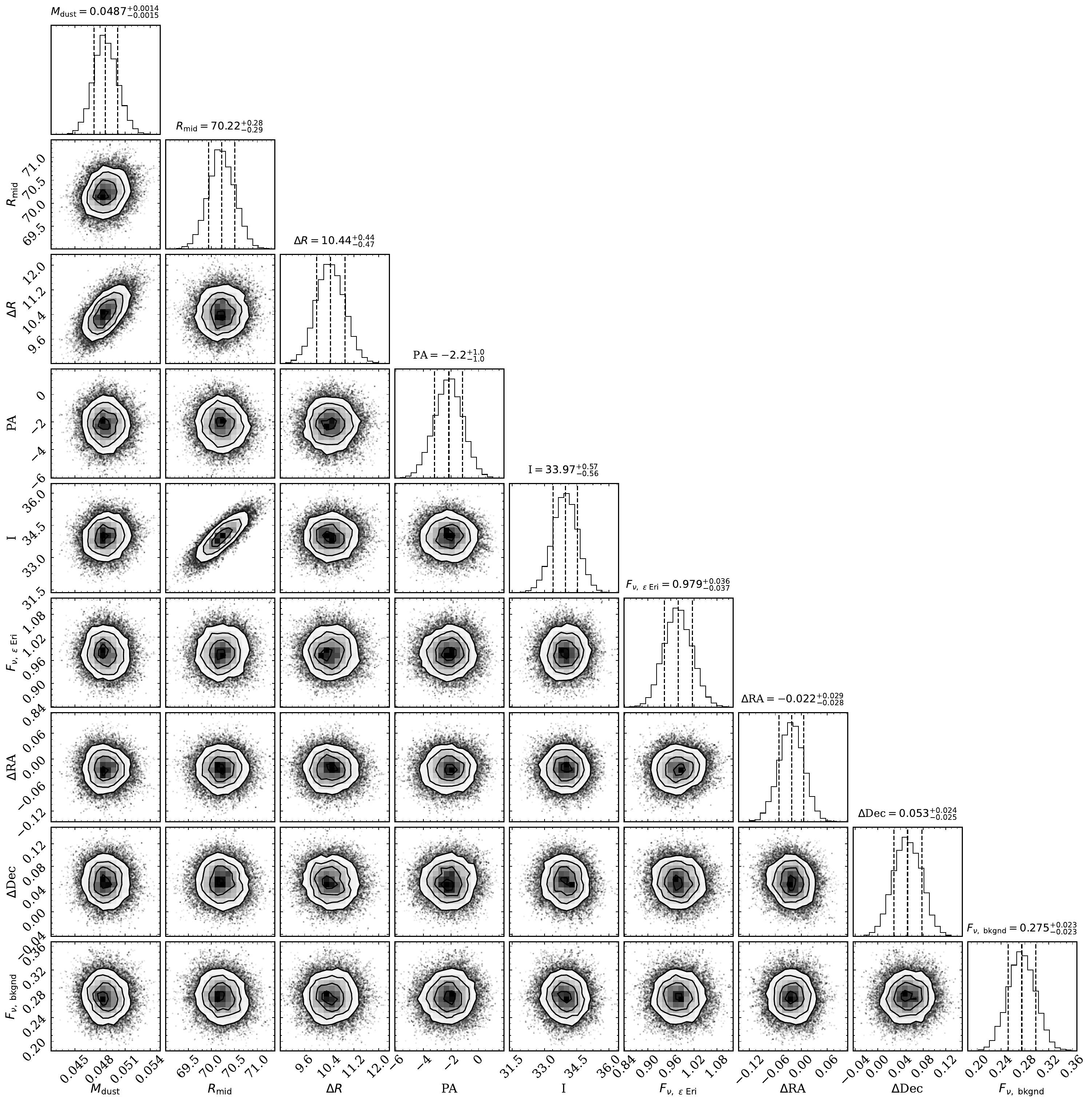}
\caption{Corner plots for the \texttt{emcee} chains for the circular model. } 
\label{fig:circular_corner}
\end{figure*}

\begin{figure*}
\includegraphics[width=\textwidth]{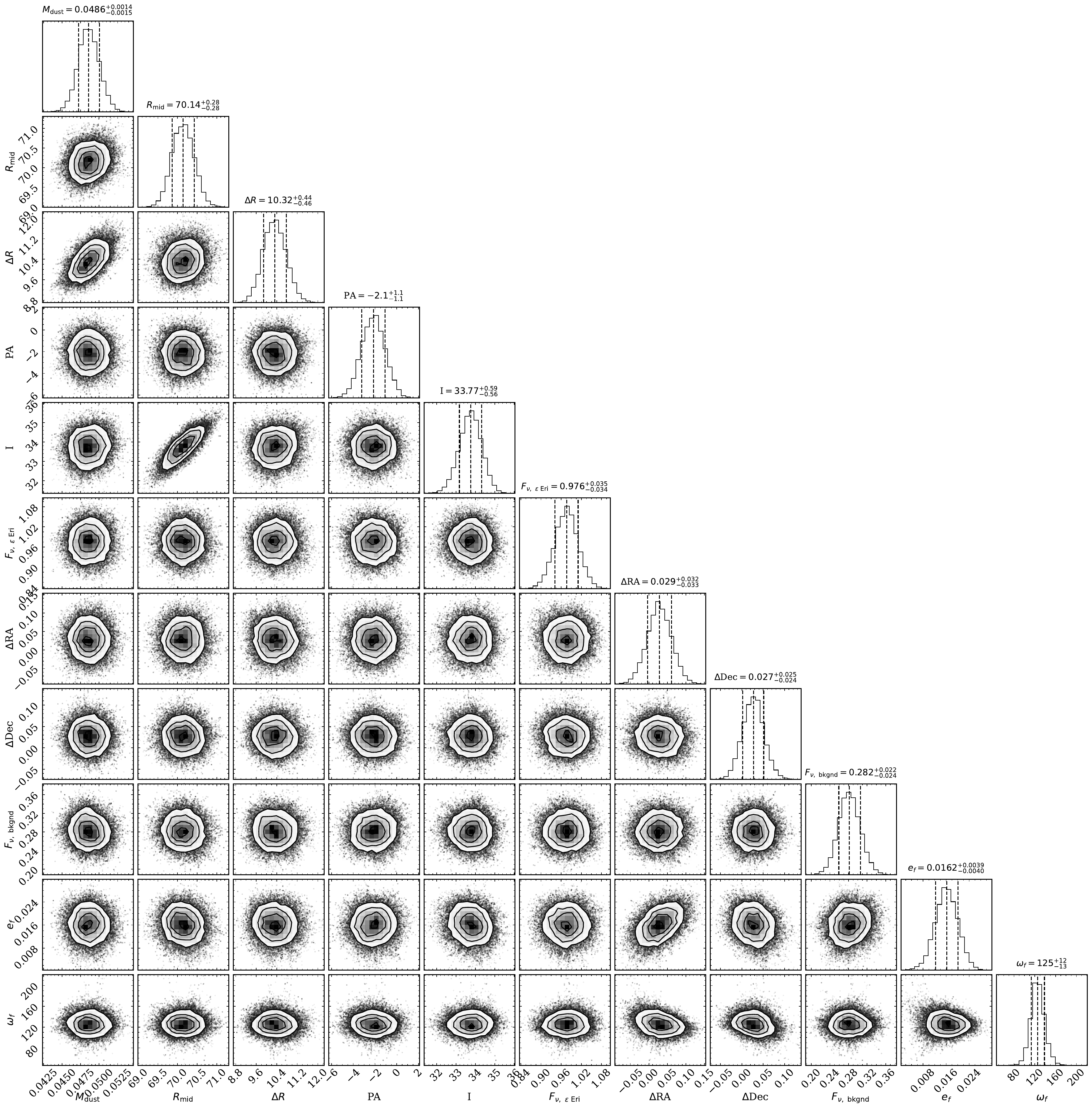}
\caption{Corner plots for the \texttt{emcee} chains for the forced-eccentric model.}
\label{fig:forced_corner}
\end{figure*}

\begin{figure*}
\includegraphics[width=\textwidth]{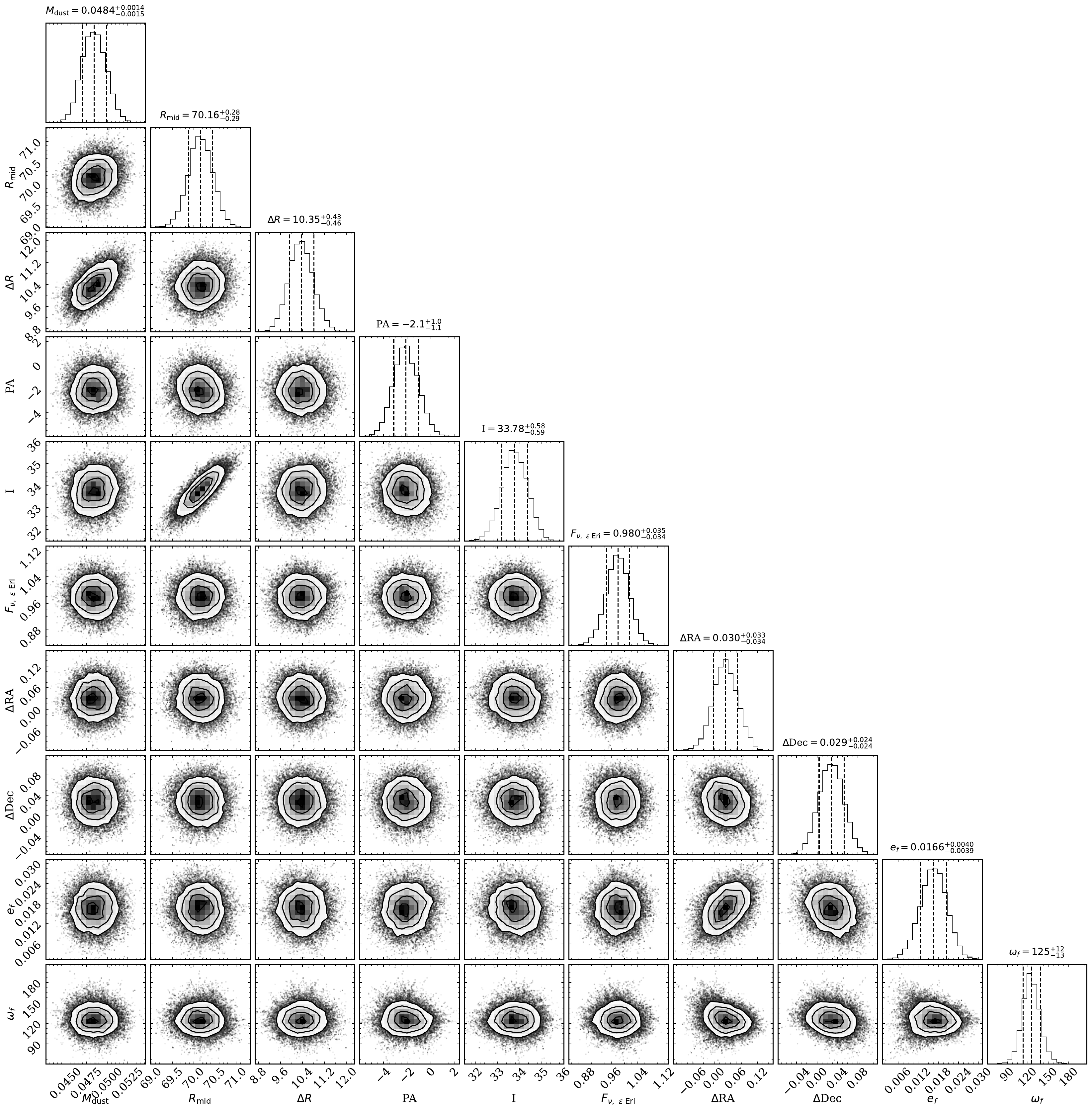}
\caption{Corner plots for the \texttt{emcee} chains for the forced eccentricity model with the background source masked by an ellipse, as discussed in \S\ref{sec:forced_ecc}.}
\label{fig:forced_corner_masked}
\end{figure*}

\begin{figure*}
\includegraphics[width=\textwidth]{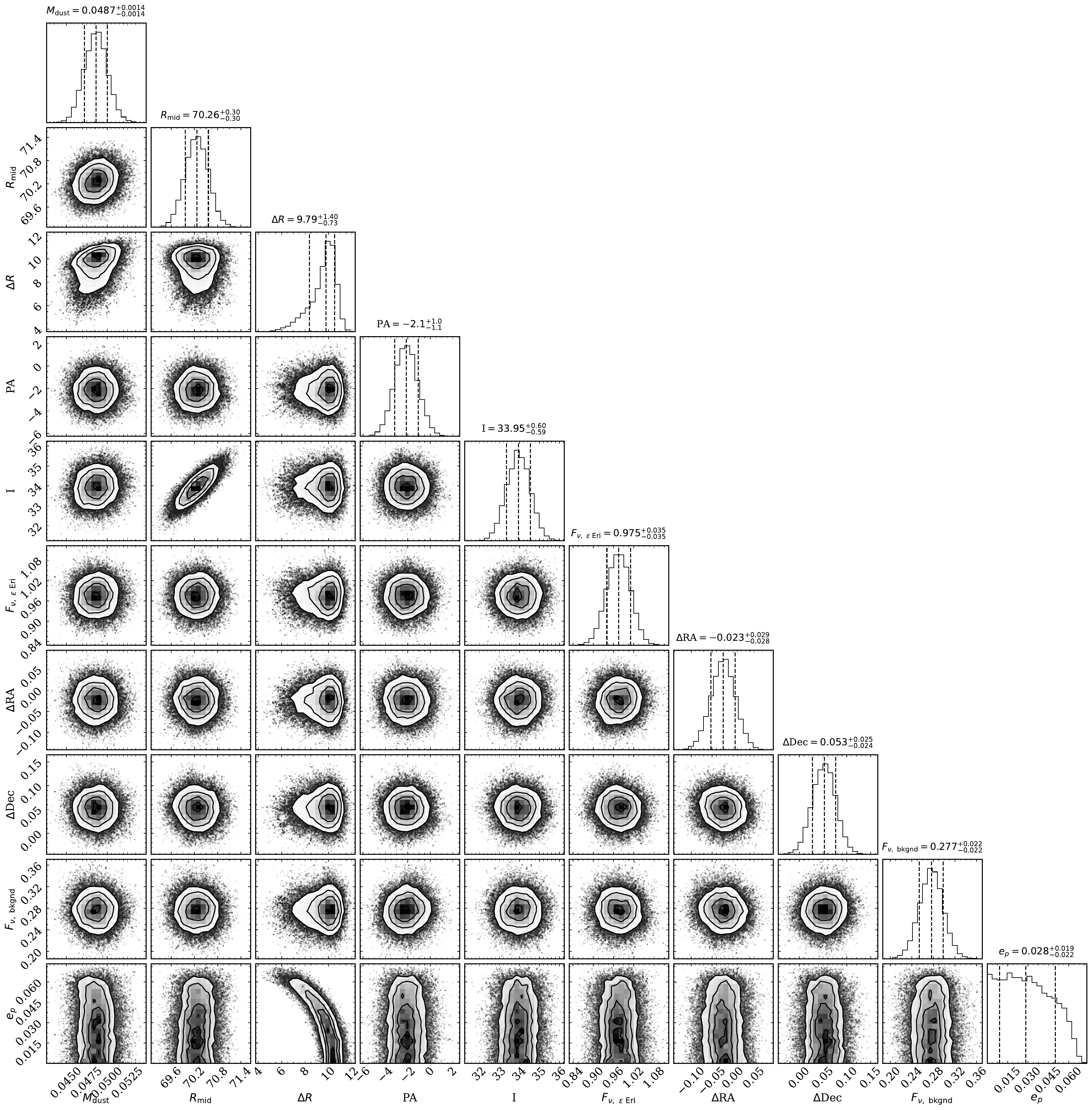}
\caption{Corner plots for the \texttt{emcee} chains for the proper-eccentric model. The expected degeneracy between the proper eccentricity and the disk width is observed, as discussed in \S\ref{sec:proper_ecc}.}
\label{fig:free_corner}
\end{figure*}

Here we present all the posterior distributions (as corner plots) based on the derived {\tt emcee} posterior distributions. 
Figures~\ref{fig:circular_corner},~\ref{fig:forced_corner}, and~\ref{fig:free_corner} show the corner plots for the Circular, Forced-eccentric and Proper-eccentric models, respectively. 
Figure~\ref{fig:forced_corner_masked} shows the corner plots for the Forced-eccentric model with the background source masked, used as a validation that the derived value of $\omega_f$ is not influenced by poorly fitting the background source.

\bibliography{epsEribiblio}{}
\bibliographystyle{aasjournalv7}
\end{document}